\documentclass[prd,twocolumn,10pt,nofootinbib]{revtex4-1}
\usepackage{hhline}
\usepackage{amsmath,amssymb}    
\usepackage{graphicx}   
\usepackage{verbatim}   
\usepackage{color}      
\usepackage{subfigure}  
\usepackage{hyperref}   
\usepackage{placeins}
 \graphicspath{{figures/}}

\newcommand{\ma}{\mu_a}
\newcommand{\msun}{M_{\odot}}
\newcommand{\vecc}[1]{\vec{\mathbf{#1}}}
\newcommand{\vechat}[1]{\hat{\mathbf{#1}}}
\newcommand{\cc}{\rm{c. c.}}
\def\l{\left}
\def\r{\right}

\newcommand{\gev}{\,\mathrm{GeV}}

\newcommand{\ev}{\,\mathrm{eV}}

\newcommand{\yr}{\,\mathrm{yr}\,}
\newcommand{\hz}{\,\mathrm{Hz}\,}
\newcommand{\kHz}{\,\mathrm{kHz}\,}
\newcommand{\kpc}{\,\mathrm{kpc}\,}
\newcommand{\mpc}{\,\mathrm{Mpc}\,}

\DeclareRobustCommand{\gobblefour}[5]{}
\newcommand*{\SkipTocEntry}{\addtocontents{toc}{\gobblefour}}

\makeatletter
\renewcommand*\l@section[2]{%
  \ifnum \c@tocdepth >\z@
  \addpenalty\@secpenalty
  \addvspace{0.5em \@plus\p@}%
  \setlength\@tempdima{1.5em}%
  \begingroup
  \parindent \z@ \rightskip \@pnumwidth
  \parfillskip -\@pnumwidth
  \leavevmode 
  \advance\leftskip\@tempdima
  \hskip -\leftskip
  #1\nobreak\hfil \nobreak\hb@xt@\@pnumwidth{\hss #2}\par
  \endgroup
  \fi}
\makeatother

\begin{document}

\title{
Discovering the QCD Axion with Black Holes and Gravitational Waves 
}

\author{Asimina Arvanitaki,}
\email{aarvanitaki@perimeterinstitute.ca}
\affiliation{Perimeter Institute for Theoretical Physics, Waterloo, Ontario, N2L 2Y5, Canada}

\author{Masha Baryakhtar,}
\email{mbaryakh@stanford.edu}

\author{Xinlu Huang,}
\email{xinluh@stanford.edu}

\affiliation{Stanford Institute for Theoretical Physics, Department of Physics,\\
  Stanford University, Stanford, CA 94305, USA}

\date{\today}

\begin{abstract}
  Advanced LIGO may be the first experiment to detect gravitational
  waves. Through superradiance of stellar black holes, it may also be
  the first experiment to discover the QCD axion with decay constant
  above the GUT scale. When an axion's Compton wavelength is
  comparable to the size of a black hole, the axion binds to the black
  hole, forming a ``gravitational atom.'' Through the superradiance
  process, the number of axions occupying the bound levels grows
  exponentially, extracting energy and angular momentum from the black
  hole. Axions transitioning between levels of the gravitational atom
  and axions annihilating to gravitons can produce observable
  gravitational wave signals. The signals are long-lasting,
  monochromatic, and can be distinguished from ordinary astrophysical
  sources. We estimate up to $\mathcal{O}(1)$ transition events at
  aLIGO for an axion between $10^{-11}$ and $10^{-10}$~eV and up to
  $10^4$ annihilation events for an axion between $10^{-13}$ and
  $10^{-11}$~eV. In the event of a null search, aLIGO can constrain
  the axion mass for a range of rapidly spinning black hole formation
  rates. Axion annihilations are also promising for much lighter
  masses at future lower-frequency gravitational wave observatories;
  the rates have large uncertainties, dominated by supermassive black hole
  spin distributions. Our projections for aLIGO are robust against perturbations
  from the black hole environment and account for our updated
  exclusion on the QCD axion of
  $6\times 10^{-13} \ev < \ma < 2 \times 10^{-11} \ev$ suggested by
  stellar black hole spin measurements.
\end{abstract}

\maketitle
\tableofcontents


\section{What is Superradiance?}
\label{sec:introduction}
A wave that scatters from a rotating black hole can exit the black
hole environment with a larger amplitude than the one with which it
came in. This amplification happens for both matter and light waves and it is called black hole
superradiance. It is an effect that has been known for nearly 50 years
\cite{Penrose:1969pc}.

Massive bosonic waves are special. They form bound states with the
black hole whose occupation number can grow exponentially
\cite{Zouros:1979iw,*detweiler,*gaina}; for fermions, Pauli's
exclusion principle makes this lasing effect impossible. This
exponential growth is understood if one considers the mass of the
boson acting as a mirror that forces the wave to confine in the black
hole's vicinity and to scatter and superradiate continuously.  This is
known as the superradiance (SR) instability for a Kerr black hole and
is an efficient method of extracting angular momentum and energy from
the black hole. Rapidly spinning astrophysical black holes thus become
a diagnostic tool for the existence of light massive bosons \cite{axiverse,Arvanitaki:2010sy}.

Black hole superradiance sounds exotic and mysterious since it naively
appears to be deeply connected with non-linear gravitational effects
in the vicinity of black holes. Instead, superradiance is a purely
kinematic effect, and black hole superradiance is just another
manifestation of the superradiance phenomenon that appears in a
variety of systems. The most famous is inertial motion superradiance,
most commonly referred to as Cherenkov radiation
\cite{Ginzburg1996}. In Cherenkov radiation, a non-accelerating
charged particle spontaneously emits radiation while moving
superluminally in a medium. The emitted radiation forms a cone with
opening angle $\cos \theta = (nv)^{-1}$, where $n$ is the index of
refraction of the medium, and radiation that scatters inside the cone
$(\omega_\gamma< \vec v \cdot \vec{k}_{\gamma})$ is amplified
\cite{Bekenstein_1998}.

Similarly, superradiance occurs for a conducting axisymmetric body
rotating at a constant angular velocity $\Omega_{\rm{cylinder}}$
\cite{zeldovich1971,*Zel_dovich_1986}. Here, superluminal motion is in
the angular direction: a rotating conducting cylinder amplifies any
light wave of the form $ e^{i m\varphi-i\omega_{\gamma} t}$ when the
rotational velocity of the cylinder is faster than the angular phase
velocity of the light:
\begin{equation}
  \label{eq:sr}
  \frac{\omega_\gamma}{m}< \Omega_{\rm{cylinder}},
\end{equation}
where $\omega_\gamma$ and $m$ are the photon energy and angular
momentum with respect to the cylinder rotation axis,
respectively. This is the same as the superradiance condition for
rotating black holes, with $\Omega_{\rm{cylinder}}$ substituted by the
angular velocity of the black hole at the horizon. The only difference
is in the (dissipative) interaction required for superradiance to occur: in the case
of a conducting cylinder, it is electromagnetism, while for black
holes, it is gravity.

Although the kinematic condition is easy to satisfy, the amplification
rate is typically small, and rotational superradiance in particular is very hard
to observe. The amplification rate is determined by the overlap of the
scattered wave with the rotating object; for non-relativistic
rotation, this overlap is proportional to $(\omega_{\gamma} R)^{2m}$
where R is the size of the object. The superradiance condition in
eq.~\eqref{eq:sr} implies that this quantity is generically much less
than $1$. As Bekenstein notes, only superradiance for the $m=1$ mode
could potentially be observed in the lab \cite{Bekenstein_1998}. For
black holes, however, several modes with $m \geq 1$ can be superradiating
within the evolution time scale of the black hole since their rotation
is relativistic.

The smallness of the superradiance rate also highlights the importance
of axisymmetry. For non-axisymmetric objects, SR modes mix
with non-SR (decaying) ones, and hence the amplification
rate is even smaller or non-existent. This is another complication for
observing rotational superradiance in the lab as well as around
astrophysical objects such as stars and planets.

To summarize, rotating black holes are just one type of system in
which superradiance can occur. However, they have special properties
that make them ideal for observing superradiance of massive bosonic
particles:
\begin{itemize}
\item{They are perfectly axisymmetric due to the no-hair theorem.}

\item{Their rotation is relativistic so the SR rate can be significant.}

\item{Gravity provides the interaction necessary for SR to
    occur, so the effect is universal for all particles.}
\end{itemize}
In particular, the superradiance rate for black holes can be
significantly faster than the dynamical black hole evolution rate. It
is maximized when the Compton wavelength of the massive bosonic
particle is comparable to the black hole size: astrophysical black
holes are sensitive detectors of bosons with masses between $10^{-20}$
and $10^{-10}$ eV.

This mass range encompasses many theoretically motivated light
bosons. In particular, the QCD axion, a pseudo-Goldstone boson
proposed to solve the strong $CP$ problem
\cite{Peccei:1977hh,*axion1,*axion2}, falls in this mass range for
high decay constant $f_a \simeq 10^{17}\gev \, \left({6 \times
    10^{-11} \ev}/{\ma}\right)$, where $\ma$ is the axion mass. Many
light axions can also arise in the landscape of string vacua \cite{axiverse}. Other classes of particles probed by superradiance
include light dilatons \cite{dilatonclock} and light gauge bosons of
hidden $U(1)$s (see \cite{intensity} and references therein). Black
hole superradiance can probe parameter space that is inaccessible to
laboratories or astrophysics since naturally light bosons have small
or no couplings to the Standard Model.

As long as the self-interaction of the boson is sufficiently weak and
its Compton wavelength comparable to the size of astrophysical black
holes, superradiance will operate, regardless of the model or the
abundance of the boson. We mostly refer to the QCD axion, but our
result is directly applicable to general scalars via $\lambda
\leftrightarrow (\ma/f_a)^2$ for a scalar with mass $\ma$ and quartic
interaction $\mathcal{L}~\supset~\lambda\phi^4/4!$. The same results
can also be approximately applied to light vector bosons.

When the superradiance effect is maximized, a macroscopic ``cloud'' of
particles forms around the black hole, giving dramatic experimental
signatures \cite{Arvanitaki:2010sy}. The signals are sizable even
after taking into account bounds suggested by measurements of rapidly
spinning black holes, which would have spun down quickly in the
presence of light bosonic particles of appropriate masses.

Black hole superradiance is fast enough to allow multiple levels
to superradiate within the dynamical evolution time scale of
astrophysical black holes. Axions occupying these levels can
annihilate to a single graviton in the presence of the black hole's
gravitational field. Levels with the same angular momentum quantum
numbers but different energies can be simultaneously populated; axions
that transition between them emit gravitational radiation. As we
will see, both axion transitions and annihilations produce
monochromatic gravitational wave radiation of appreciable intensity.
The gravitational wave frequency and strain for GUT- to Planck-scale
QCD axions fall in the optimal sensitivity band for Advanced LIGO
(aLIGO) \cite{Harry:2010zz} and VIRGO \cite{Accadia:2011zzc}.

The annihilations signature is also promising at future, low-frequency
gravitational wave observatories. Another signal relevant for
bosons with self-coupling stronger than the QCD axion is the
``bosenova'' effect \cite{Arvanitaki:2010sy}, where the bosonic cloud
collapses under its self-interactions, producing periodic gravitational
wave bursts.

In this paper, we focus on the prospects for detecting gravitational
wave signals at aLIGO and discuss the reach for future gravitational wave detectors
operating at lower frequencies. In section~\ref{sec:theory}, we review
the parameters for black hole superradiance and how it evolves for an
astrophysical black hole. In section~\ref{sec:gw_signals}, we estimate
expected event rates at aLIGO and at future lower frequency
detectors. In section~\ref{sec:spin_limit}, we revisit bounds from
black hole spin measurements and include our results for both stellar
and supermassive black holes. We examine the effects of black hole
companion stars and accretion disks on superradiance in
section~\ref{sec:environment}, and conclude in section VI.

\section{Theoretical Background}
\label{sec:theory}

\subsection{The Gravitational Atom in the Sky}
\label{sec:sr_theory}
The bound states of a massive boson with the black hole (BH) are
closely approximated by hydrogen wave functions: except in very
close proximity to the black hole, the gravitational potential is
$\propto 1/r$. The ``fine-structure constant'' $\alpha$ of the
gravitational atom is:
\begin{equation}
  \alpha = r_g \ma, \qquad r_g \equiv G_N M, 
\end{equation}
where $r_g$ is the gravitational radius of the BH, $M$ its mass, and
$\ma$ the boson's mass. Throughout this paper, we use units where $ c
= \hbar = 1$. Like the hydrogen atom, the orbitals around
the black hole are indexed by the principal, orbital, and magnetic
quantum numbers $\{n, \ell, m\}$ with energies:
\begin{equation}
  \label{eq:energies}
  \omega \simeq \ma \left(1-\frac{\alpha^2}{2{n}^2}\right).
\end{equation}
The orbital velocity is approximately $ v\sim \alpha/\ell$, and the
axions form a ``cloud'' with average distance
\begin{equation}
  \label{eq:r_c}
  r_c \sim \frac{{n}^2}{\alpha^2}r_g
\end{equation}
from the black hole.

A level with energy $\omega$ and magnetic quantum number $m$ can
extract energy and angular momentum from the black hole if it
satisfies the superradiance condition analogous to
eq.~\eqref{eq:sr}:
\begin{equation}
  \label{eq:sr_cond}
  \frac{\omega}{m} < \omega^+, \quad \omega^+ \equiv \frac{1}{2}\left(\frac{a_*}{1+\sqrt{1-a_*^2}}\right)r_g^{-1},
\end{equation}
where $\omega^+$ can be thought of as the angular velocity of the
black hole and $0 \leq |a_*| < 1$ is the black hole spin ($a_*\equiv
a/r_g$ in Boyer-Lindquist coordinates). The SR
condition requires
\begin{equation}
  \label{eq:sr_br}
  \alpha/\ell \leq 1/2,
\end{equation}
with the upper bound saturated for $m = \ell$ and extremal black holes
($a_* = 1$), so superradiating bound states are indeed
well-approximated by solutions to a $1/r$ gravitational potential
($r_c \gg r_g$) with sub-leading relativistic corrections ($v^2 \ll
1$).

The occupation number\footnote{The axion cloud surrounding the BH is
  described by a classical field, and therefore does not have a
  well-defined occupation number $N$. In this paper we define the
  occupation number as the average value of bosons in the cloud. }
$N$ of levels that satisfy the SR condition grows exponentially with a
rate $ \Gamma_{\text{sr}}$,
\begin{align}
  \left.\frac{dN}{dt}\right|_{\text{sr}} &=
  \Gamma_{\text{sr}}N,\\
  \Gamma_{\text{sr}}^{n\ell m}(a_*,\alpha, r_g) &=\mathcal{O}(10^{-7}\textrm{--}10^{-14})\, r_g^{-1}.\nonumber
\end{align} 
The boson is not required to be dark matter or be physically present
in the vicinity of the black hole: just like spontaneous emission,
superradiance can start by a quantum mechanical fluctuation from the
vacuum, and proceed to grow exponentially. If the SR condition is
satisfied, the growth will occur as long as the rate is faster than
the evolution timescales of the BH, the most relevant of which is the
Eddington accretion time, $\tau_{\mathrm{Eddington}} = 4\times 10^{8}$
years. The growth stops when enough angular momentum has been
extracted so that the superradiance condition is no longer
satisfied. At that point the number of bosons occupying the level is
\begin{equation}
  \label{eq:nmax}
  \!\!N_{\rm{max}}\simeq\frac{G_NM^2}{m} \Delta a_* \sim 10^{76} \left(\frac{\Delta a_* }{0.1}\right)\!\!\left(\frac{M}{10\msun}\right)^2,
\end{equation}
where $\Delta a_* = \mathcal{O}(0.1)$ is the difference between the
initial and final BH spin. 
\begin{figure}[t]
  \centering
  \includegraphics[width=\linewidth]{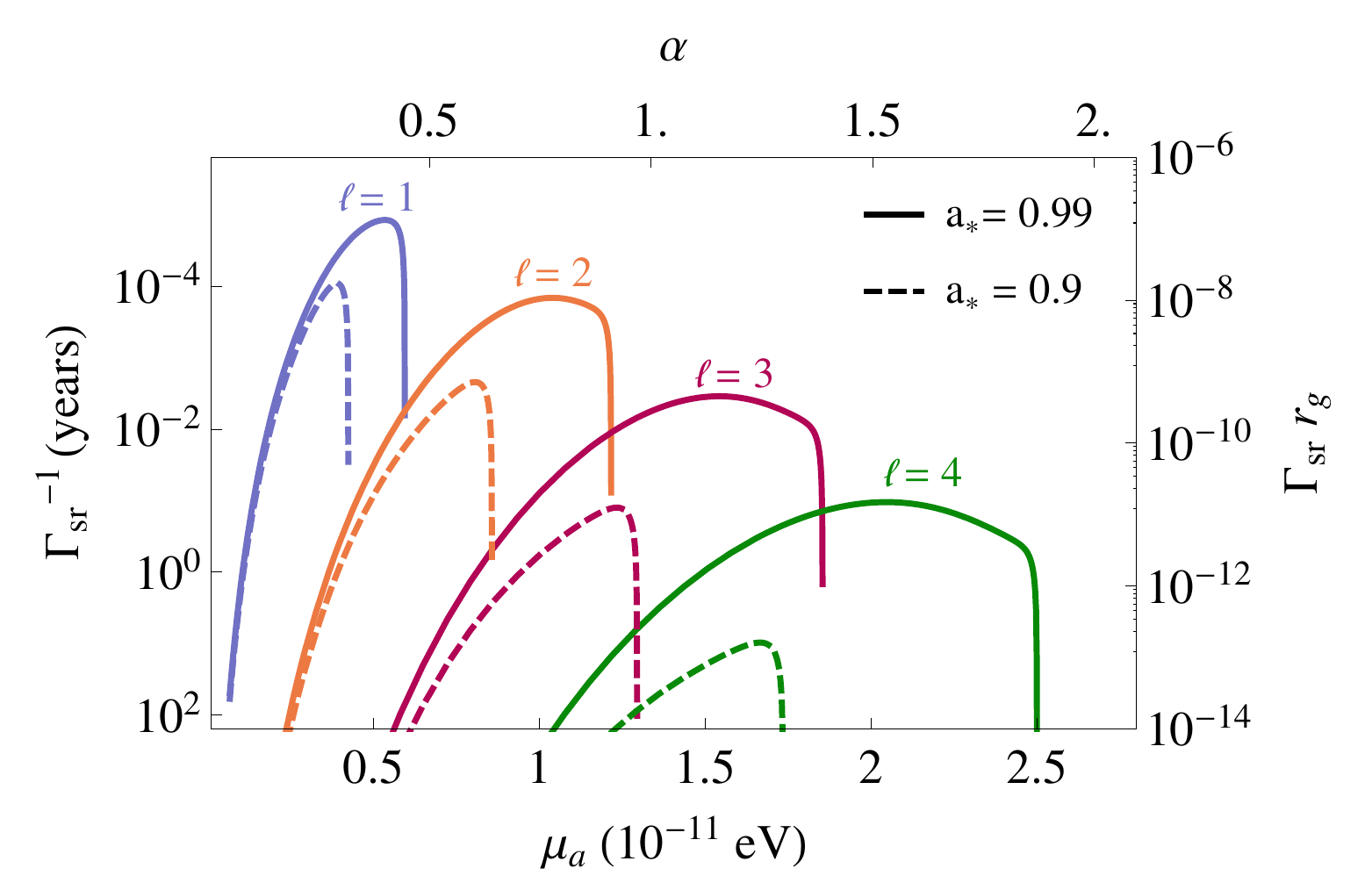}
  \vspace{-6.5mm}
  \caption{Superradiance times of levels $\ell = 1$ to $4$ (left to
    right) for spins $a_* = 0.99$ and $0.90$, fixing $m=\ell$ and $n =
    \ell + 1$. Time in years is shown for a $10\, M_{\odot}$ black
    hole as a function of boson mass $\mu_a$; on the right axis, we
    show the dimensionless superradiance rate $\Gamma_{\rm{sr}} r_g$
    as a function of the gravitational coupling $\alpha$ (top axis).}
  \label{fig:sr_rates}
\end{figure}

The superradiance rates (or dumping rates for the levels that are not
superradiating) are given by the small imaginary part of the energy of
a free-field solution in the Kerr background. Unless otherwise
specified, we use the semi-analytic approach for massive spin-0 fields
presented in \cite{Arvanitaki:2010sy}, which agrees well with
analytical formulae for $\alpha/\ell~\ll~1$ \cite{detweiler} and the
WKB approximation for $\alpha/\ell~\sim~\mathcal{O}(1/2)$
\cite{Zouros:1979iw,Arvanitaki:2010sy}, as well as with partial
numerical results in \cite{dolan}. Rates for massive spin-1 fields are
expected to be larger, and some numerical progress has been made
toward calculating them \cite{Pani:2012bp,*Rosa:2011my}; we choose to
focus on the spin-0 case (including the QCD axion) for
the remainder of this paper, but further studies with spin-1 fields
are very worthwhile.

In fig.~\ref{fig:sr_rates} we show representative values of the
superradiance rates, $\Gamma_{\text{sr}}$, and we list sample values
in Table \ref{tab:timescales} along with typical BH evolution time
scales. The rate varies with the relevant parameters of the system as
follows:
\begin{itemize}
\item $r_g$ --- The dimensionless quantity $\Gamma_{\rm{sr}} r_g$
  depends only on the coupling $\alpha$, BH spin $a_*$ and the quantum
  numbers of the state; the physical SR time can be as
  short as $100$ s for stellar black holes and is longer for heavier
  black holes.
\item \textbf{$\alpha$} --- For given level, $\Gamma_{\rm{sr}}$ is a
  steep function of the coupling, reaching its maximum close to the
  SR boundary.  A single BH is sensitive to a range of boson
  masses: stellar BHs ($2$--$100 M_{\odot}$) correspond to masses of
  $10^{-13}$--$10^{-10}$ eV, and supermassive BHs ($10^6$--$10^8
  M_{\odot}$) to masses of $10^{-19}$--$10^{-16}\ev$.
\item \textbf{$a_*$} --- The dependence of $\Gamma_{\rm{sr}}$ on spin
  enters primarily through the SR condition. The upper bound on
  $\alpha/\ell$ becomes smaller than $1/2$ for lower spin BHs, and the
  maximum SR rate is thus smaller than for equivalent BHs
  with higher spins.

\item $\ell$ ---  $\Gamma_{\rm{sr}}$ decreases with increasing $\ell$,
  and the dependence is strong: for
  $\alpha/\ell \ll 1$, $\Gamma_{\rm{sr}} \propto \alpha^{4\ell}$ \cite{detweiler}, while
  for $\alpha/\ell \sim 0.5$, the WKB approximation \cite{Zouros:1979iw,Arvanitaki:2010sy} gives $\Gamma_{\rm{sr}}
  \propto e^{-3.7\alpha}=(0.15)^\ell$. 
\item $m$ ---  $\Gamma_{\rm{sr}}$ is largest for $m=\ell$ and is
  much smaller for $m < \ell$. Unless otherwise specified, we
  only consider levels with $m=\ell$ below.
\item $n$ --- For fixed $\ell$ and $m$, the dependence on $n$ is mild and
  $\Gamma_{\rm{sr}}$ generally decreases with larger $n$.
\end{itemize}

\begin{table*}[t!]
  \begin{center}

    \begin{tabular}{  c c  c | c |  c  }
      \hline
      Process & & (see also)  & \,Stellar BHs\, & \,Supermassive BHs\,\\ 
      \hline\hline
      Superradiance ($2p, \alpha = 0.3, a_* = 0.9$) &
      $\Gamma_{\rm{sr}}^{-1} $ & fig.~\ref{fig:sr_rates}& $10^{-4} \yr$& $100 \yr$\\ 
      Superradiance ($5g, \alpha = 1.2, a_* =
      0.9$)&$\Gamma_{\rm{sr}}^{-1}$& fig.~\ref{fig:sr_rates}& $10 \yr$ & $10^7 \yr$\\ 
      Regge trajectory ($2p, \alpha = 0.3, f_a=10^{17}\gev$)
      &$\tau_{\rm{regge}}$ & eq.~\eqref{eq:regge_time} & $10^{6} \yr$ & $10^{12} \yr$\\ 
      Eddington Accretion, $(\dot{M}/M)^{-1}$ &  $\tau_{\rm{eddington}}$ & & $4 \times 10^8$ yr&  $4 \times 10^8$ yr\\
      \hline
    \end{tabular}
    \caption{Characteristic superradiance timescales. We use
      $10 \,M_{\odot}$ and $10^{7} \,M_{\odot}$ as representative 
      stellar and supermassive black hole masses. The $2p$ level is the
      most relevant for annihilation signals, and the $5g$ level for transition signals.}
    \label{tab:timescales}
  \end{center}
\end{table*}

So far, we have considered a free bosonic field. Self-interactions
between bosons will affect superradiance when the interaction energy
becomes comparable to the binding energy of the boson in the cloud
\cite{Arvanitaki:2010sy}. Axions, for example, have attractive
self-interactions which cause the cloud to collapse when it reaches a
critical size; laboratory experiments have observed such collapse of
bosonic states, known as ``bosenova", in the analogous system of
trapped Bose-Einstein condensates \cite{bec}.  Bosenovae can even be a portal into a hidden
sector to which a generalized light axion couples
\cite{Dubovsky:2010je}. After a bosenova occurs, the superradiant
growth restarts.  Even weak self-interactions can have a significant
effect: for example, for an axion with decay constant $f_a$, the
critical \mbox{bosenova occupation} number is
\begin{equation}
  \label{eq:nbosenova}
  N_{\rm{bosenova}} \simeq
  10^{78}\,c_0\,\frac{n^4}{\alpha^3} \left(\frac{M}{10\msun}\right)^2 \left(\frac{f_a}{M_{\rm{pl}}}\right)^2,
\end{equation}
where $M_{\rm{pl}} = 2 \times 10^{18} \gev$ and $c_0\sim 5$ is determined by numerical simulation
\cite{Yoshino:2012kn}.  Comparing eqs.~\eqref{eq:nmax} and
\eqref{eq:nbosenova}, we see that the bosenova occurs before all the
spin is extracted when
\begin{equation}
  \label{eq:fa_bosenova}
  f_a \lesssim 2\times 10^{16} \gev\,\frac{1}{\sqrt{n}}
  \left(\frac{\alpha/n}{0.4}\right)^{\frac{3}{2}}\!\!\left(\frac{\Delta a_{*}}{0.1}\right)^{\frac{1}{2}}\left(\frac{5}{c_0}\right)^{\frac{1}{2}}.
\end{equation}
For the QCD axion this gives $\ma>3\times 10^{-10}\ev$, too heavy to
be relevant for astrophysical black holes ($M \gtrsim 3
\msun$). Nevertheless, the bosenova can lead to interesting
gravitational wave signals for axion-like particles and other light
bosons (section~\ref{sec:bosenova}). For strongly interacting bosons,
the superradiance instability can be slowed to a stand-still with the
cloud collapsing before it can grow to macroscopic size.

In this section, we use the results of
\cite{Zouros:1979iw,*detweiler,*gaina,Arvanitaki:2010sy};
we refer readers to these references for further details. For a broad
review of SR, see \cite{Brito:2015oca}.

\subsection{A (Not So) Brief History of Superradiance}
\label{sec:regge}

\begin{figure}[t]
  \centering
  \includegraphics[width=\linewidth]{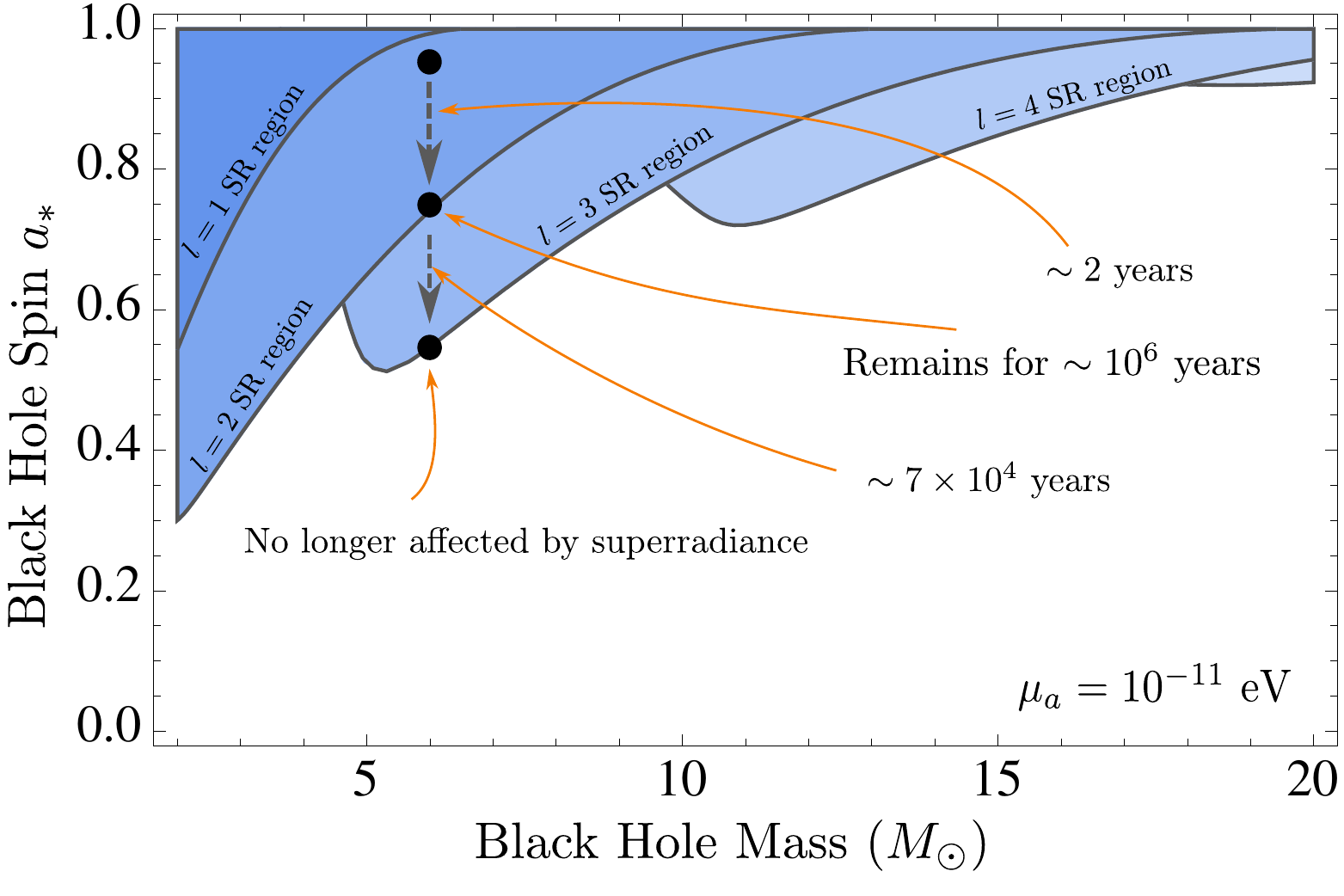}
  \caption{Effect of superradiance for a QCD axion with mass $\mu_a =
    10^{-11}\ev$ and decay constant $f_a = 6\times
    10^{17}\gev$. Shaded regions correspond to BH parameters
    which would result in spin down within a binary lifetime ($10^6$
    years), for $\ell=1$ (dark blue) to $\ell=5$ (light blue)
    levels. We also show an example evolution of a $6\msun$ black hole
    with initial spin $a_* = 0.95$. }
  \label{fig:regge}
\end{figure}

The superradiance condition can be satisfied for several levels of the
black hole-axion ``atom'', and for each level and boson mass, there is
a region in the BH spin vs. mass plane that is affected
(fig.~\ref{fig:regge}). As discussed previously, the superradiance
condition is a kinematic one and SR can affect BHs with masses a
factor of $10$ to $100$ around the optimal value. The affected region is
set by the SR condition and is further limited by whether
superradiance happens faster than the accretion rate of the BH.

In order to understand how superradiance affects astrophysical black
holes, let us assume there exists a boson with mass $\ma=10^{-11}$ eV
and self-interaction strength of the QCD axion (decay constant
$f_a =6\times 10^{17}\gev$). The Compton wavelength of this particle is
$20$~km, the size of a typical stellar BH horizon.

Consider a BH that is born with spin $a_*= 0.95$ and mass $6
M_{\odot}$. Once the environment settles to a steady state after the
supernova explosion, superradiant levels begin to grow exponentially
with their respective SR rates. The fastest-growing level
dominates --- in this case, the $\ell=2$ level, since the smallest $\ell$
that satisfies the SR condition has the largest rate. It takes $\log
N_{\rm{max}}\sim 200$ e-folds --- in this case, about $2$~years --- of
growth to extract enough spin so that the SR condition is no longer
satisfied for the $\ell=2$ level. While losing $20\%$ of the spin, the
BH only loses about $5\%$ of its mass, because the cloud is larger in
extent than the black hole and so more efficient at carrying angular
momentum \cite{Arvanitaki:2010sy}. As the cloud grows, the gravitational wave signal from axion
annihilation (section \ref{sec:annihilation}) increases until
reaching a maximum when the SR condition is no longer satisfied.

At this point, we expect superradiance to start populating the $\ell=3$ at a
slower rate.  If self-interactions are present, this does not happen
right away: the $\ell=2$ level perturbs the potential around the black
hole such that the $\ell=3$ level mixes with levels that do not
satisfy the SR condition. Therefore the $\ell = 3$ level does not grow until the
$\ell=2$ level is depleted to the point when level mixing is a
subdominant effect. The time scale for the $\ell$th level to be
depleted sufficiently is dominated by two-boson to one
graviton annihilations \cite{Arvanitaki:2010sy},
\begin{equation}
  \label{eq:regge_time}
  \tau_{\rm{regge}} \simeq
  (N_{\rm{bosenova}}\Gamma_a)^{-1}\l|{\Gamma^{\ell-1}_{\mathrm{sr}}}/{\Gamma^{\ell+1}_{\mathrm{sr}}}\r|^{1/2},
\end{equation}
where $\Gamma_a$ is the annihilation rate
(section~\ref{sec:annihilation}).
For interacting particles $\tau_{\rm{regge}}$ can be much
longer than the superradiance time since $(N_{\rm{bosenova}}\Gamma_a)< \Gamma_{\rm{sr}}$ and
$\l|{\Gamma^{\ell-1}_{\mathrm{sr}}}/{\Gamma^{\ell+1}_{\mathrm{sr}}}\r|^{1/2}\gg
1$. The black hole can therefore spend a long
time on the line where $\omega= m \omega_+$. This line thus
defines for black holes the analog of Regge trajectories in particle
physics. If black hole spin measurements become accurate enough, we
could diagnose the presence of an axion by fitting the curve of BH
spin vs. mass to the superradiance condition.

Once the $\ell =2$ level is depleted through annihilations, the
$\ell=3$ level starts to grow and the BH makes another jump in the BH
spin vs. mass plane. The previous process then repeats itself, but for
the parameters chosen, the $\ell=4$ superradiance rate is too slow 
and once the spin drops to $a_* = 0.55$, superradiance no longer
affects the BH.

If the black hole is heavier such that the $\ell = 4$ superradiance
rate is significant, the $5g$ and $6g$ levels grow with comparable
superradiance rates to large occupation numbers.  This sets the stage
for a large gravitational wave signal from level transitions
(section~\ref{sec:transition}); $\ell =4$ is the smallest $\ell$ for
which significant transition signals occur.

The BH trajectory is more complicated if the bosenova is possible. In
that case, the cloud reaches a maximum size of $N_{\rm{bosenova}}$ and
collapses before saturating the superradiance condition. Then, the
bosenova has to repeat many times before the superradiance condition
is saturated, and the occupation number of the cloud at the Regge
trajectory is smaller by a corresponding factor. As we discuss in
section~\ref{sec:bosenova}, the periodic repetition of bosenovae
can give rise to interesting signals.

For supermassive black holes the story changes slightly, since their
spin and mass are acquired through accretion. As a supermassive BH
grows, spin extraction by the cloud happens adiabatically with 
black hole accretion, moving the BH along the boundary of the region in the spin
vs. mass plane affected by superradiance. Only a violent event such as
a merger will perturb the system enough so the BH can jump between
different levels.  The long time spent on the trajectories can lead to
exciting annihilation signals at low-frequency gravitational wave
detectors (section~\ref{sec:annihilation_future}).

\section{Gravitational Wave Signals}
\label{sec:gw_signals}

Processes that have forbiddingly small rates for a single particle can
be enhanced in the bosonic cloud, since the occupation number of a
single level in the BH-boson gravitational atom can be exponentially
large. Transitions of bosons between two different levels are enhanced
by $N_1 N_2$ where $N_i$ is the occupation number for each level;
two-boson annihilations to a single graviton are enhanced by
$N_i^2$. Because of this ``lasing" effect, the peak strain of the
resulting gravitational waves (GW) can be within reach of GW
detectors. Superradiance for stellar black holes can lead upcoming
observatories, Advanced LIGO \cite{Harry:2010zz} and Advanced VIRGO
\cite{Accadia:2011zzc} --- beginning science runs in 2015-2016
\cite{Aasi:2013wya} --- into the realm of discovery. Superradiance
for supermassive black holes has exciting prospects for future,
low-frequency observatories.

\begin{table*}[t!]

  \begin{center}

    \begin{tabular}{  c c  c | c |  c  }
      \hline
      & & (see also)  & Stellar BHs & Supermassive BHs\\ 
      \hline\hline
      Transition ($6g\rightarrow 5g, \alpha = 1.2$) & $\Gamma_t^{-1}$ & eq. \eqref{eq:transition_rate}&
      $10^{72} \yr $ & $10^{90}\yr  $ \\
      Annihilation ($2p$, $\alpha = 0.3$)& $\Gamma_a^{-1}$ & eq. \eqref{eq:annih_rate}& $10^{79}\yr$&
      $10^{97} \yr$\\
      \hline
      Maximum number of axions in the cloud & $N_{\rm{max}}$ & eq. \eqref{eq:nmax}& $10^{76}$ & $10^{88}$\\
      \hline
      Transition signal length ($6g\rightarrow 5g,\alpha = 1.2$) \,\,\,\,& $\mathcal{O}(1) \times
      \Gamma_{\rm{sr}}^{-1}$ \,\,\,\,& fig.~\ref{fig:two-level-sys}& $5$ yr  & $5\times 10^6$ yr\\
      Annihilation signal length ($2p, \alpha = 0.3$)
      & $(N_{\rm{max}}\Gamma_a)^{-1}$&eq.~\eqref{eq:annihtime}& $10^3$ yr& $10^9$ yr\\
      \hline

    \end{tabular}
    \caption{Characteristic GW signal timescales and parameters. We use $10
      M_{\odot}$ and $10^{7} M_{\odot}$ as representative stellar and
      supermassive black hole masses, and spin of $ a_* = 0.9$.  Signal length is
      defined as the duration for which signal is larger than $1/e$ of 
      its maximum (see sections
      \ref{sec:transition} and \ref{sec:annihilation}).}
    \label{tab:timescales2}

  \end{center}

\end{table*}

There are three types of GW signals from the bosonic cloud:
\begin{itemize}
\item graviton emission from level transitions
\item axion annihilations into gravitons 
\item bosenova collapse of the axion cloud
\end{itemize}
The axions involved in transitions and annihilations are in exact
energy eigenstates of the black hole potential and thus emit
monochromatic GWs.\footnote{This disagrees with what was stated in
  \cite{Arvanitaki:2010sy} regarding the monochromaticity of GWs from
  annihilations. We thank S. Dimopoulos and S.  Dubovsky for
  discussions clarifying this issue.} As the occupation number of a
level changes, the axion energy receives a correction $\Delta E \sim
E_{\rm{bind}} N/(2N_{\rm{bosenova}})$ due to axion self-interactions
\cite{Arvanitaki:2010sy}, leading to a small frequency drift of the
emitted GW which can be used to distinguish the signal from
astrophysical sources.

In this section, we compute the experimental reach of GW observatories
and estimate expected event rates.  Table \ref{tab:timescales2}
summarizes the timescales typical for these processes. We focus on
light axions as a prime example of bosons relevant for superradiance,
since their mass and self-interaction are naturally small due to shift
symmetry. When relevant, we assume the self-coupling is that of the
QCD axion in this section.

To calculate the GW strain we use
\begin{equation}
  \label{eq:0}
  h = \left(\frac{4 G_N P}{r^2 \omega^2}\right)^{1/2}
\end{equation}
for a source emitting power $P$ of angular frequency $\omega$ at a
distance $r$ away from the Earth. We do not include the effects from the
angular dependence and orientation of the GW detectors.

We compare the signal strain to GW detector sensitivity $h_{\rm{det}}$ for a
search with $N_{\mathrm{seg}}$ segments of $T_{\mathrm{coh}}$ coherent
integration times:
\begin{equation}
  \label{eq:16}
  h_{\rm{det}}= n(\sigma) C_{\rm{tf}}\frac{\sqrt{S_h}}{N_{\mathrm{seg}}^{1/4}T_{\mathrm{coh}}^{1/2}},
\end{equation}
where $S_h$ is the detector noise spectral density, $n(\sigma)$ is the
signal to noise ratio for a desired signal significance, and
$C_{\rm{tf}}$ the trials factor. We use $C_{\rm{tf}} = 10$ in
this section as a realistic value since we expect frequency drifts to
be unimportant for this search \cite{privatecomm} (compared to
$C_{\rm{tf}} \sim 20$ in the current LIGO periodic gravitational wave
search, where $H\simeq n(\sigma)\,{N_{\mathrm{seg}}^{1/4}} C_{\rm{tf}}\sim
150$ \cite{aasi2013einstein}).

\subsection{Level Transitions}
\label{sec:transition}

Analogously to atomic transitions emitting photons, level
transitions of axions around black holes emit gravitons. The
GW angular frequency for transitions between an ``excited" and
a ``ground" level with principal quantum numbers ${n}_e$ and ${n}_g$,
respectively, is
\begin{equation}
  \omega_{\rm{tr}} = \frac{1}{2}\ma \alpha^2\left(\frac{1}{{n}_g^2} - \frac{1}{{n}_e^2}\right).
\end{equation}

When the two levels dominate the SR evolution, their occupation
numbers $N_{e,g}$ evolve with their respective SR rates, modified by
axions transitioning from the excited to the ground state via graviton
emission,
\begin{equation}
  \label{eq:two_level_dynamics}
  \frac{d N_g}{d t} = \Gamma^{\rm{sr}}_{g}N_g + \Gamma_{t}N_g N_e,\,\,\,\,
  \frac{d N_e}{d t} = \Gamma^{\rm{sr}}_{e}N_e - \Gamma_{t}N_e N_g,
\end{equation}
where $\Gamma^{\rm{sr}}_{g,e}$ are the superradiance rates for the two
levels and $\Gamma_t$ the transition rate for a single axion. A
quadrupole formula estimate gives
\begin{equation}
  \label{eq:transition_rate}
  \Gamma_{t} \sim \frac{2 G_{N}\omega^5}{5} \ma^2r_c^4 =
  \mathcal{O}(10^{-6}-10^{-8}) \frac{G_{N}\alpha^{9}}{r_g^3}.
\end{equation}
For our numerical results, we compute more precise rates (see
app.~\ref{sec:app_GW_power}).

\begin{figure}[t]
  \hspace{-5mm}
  \vspace{-3.5mm}
  \subfigure{\includegraphics[trim = 3mm 5mm 13mm 5mm, clip, width=.5\linewidth]{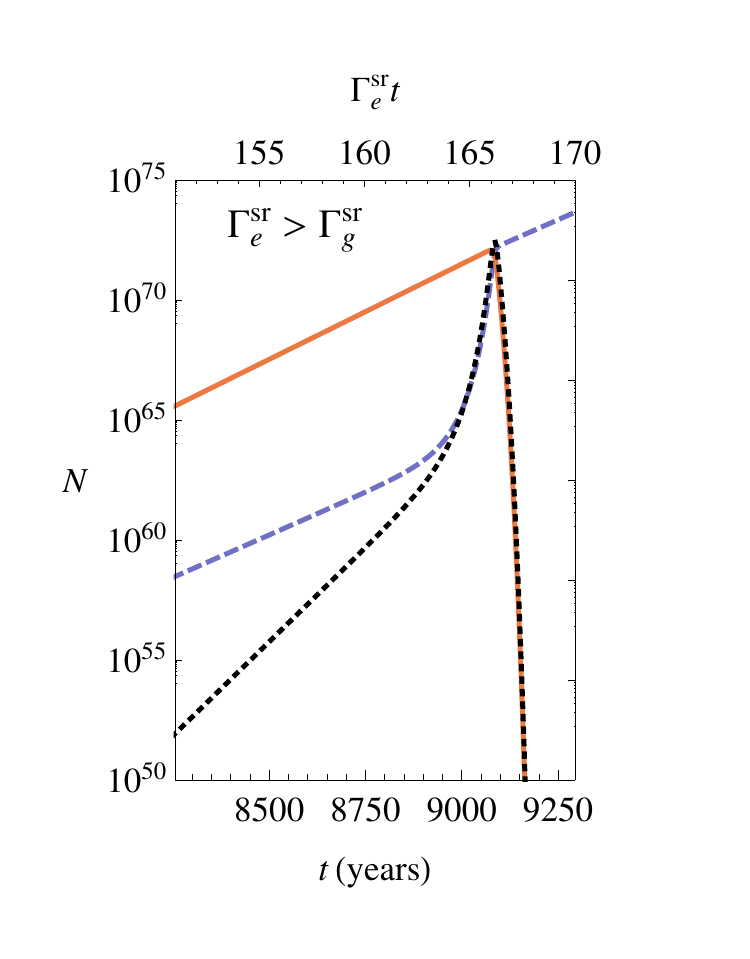}}  
  \subfigure{\includegraphics[trim = 13mm 5mm 3mm 5mm, clip, width=.5\linewidth]{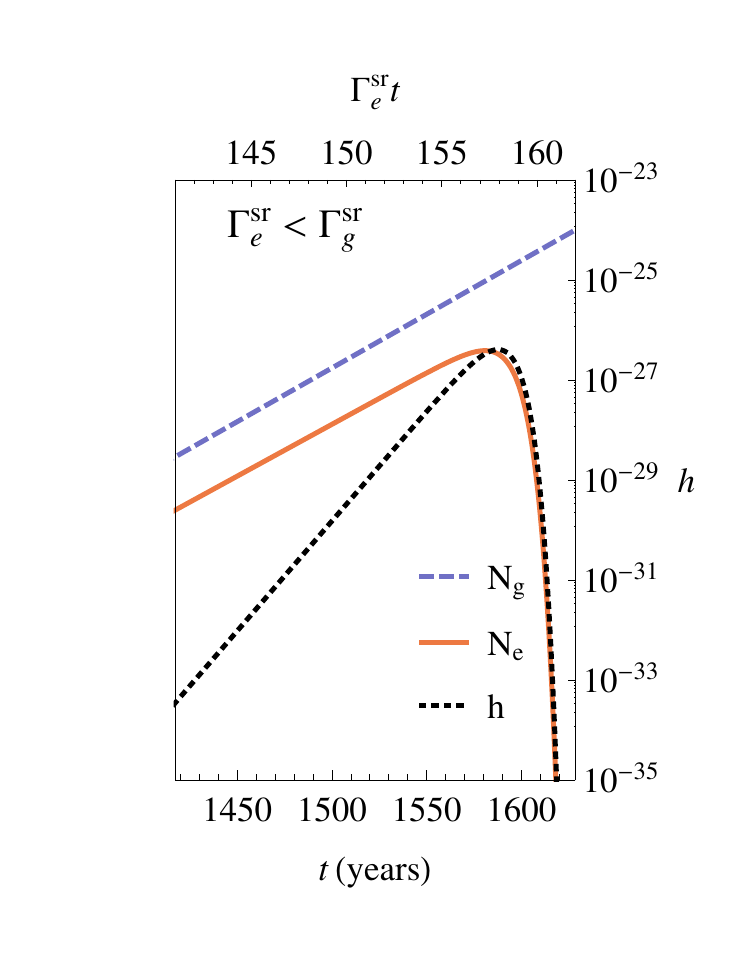}}
  \caption{Time evolution of ground and excited levels' occupation
    numbers (left $y$-axis) and the resulting gravitational wave
    signal strain (right $y$-axis) for the $6g\rightarrow 5g$
    transition around a $10\,\msun$ black hole with spin $a_* = 0.9$, 
    10 kpc away. The peak signal is larger when $\Gamma_e > \Gamma_g$
    (left, $\alpha = 1$) than the case $\Gamma_e < \Gamma_g$ (right,
    $\alpha = 1.25$). The initial occupation numbers of both levels
    are set to 1 when $t = 0$, and while the time in years differs
    significantly, the characteristic timescales for the signals are
    set by the superradiance rates (top axes) in both cases. }
  \label{fig:two-level-sys}
\end{figure}

Although the single axion transition rate is tiny ($\Gamma_{t}\lll
\Gamma^{\rm{sr}}_{e, g}$), the emission of gravitational waves is enhanced
by the occupation numbers of each level:
\begin{equation}
  h_{\rm{tr}}(t)= \sqrt{\frac{4 G_N}{r^2\omega_{\rm{tr}}}\Gamma_tN_g(t)N_e(t)}.
\end{equation}
When the axion clouds are small, transitions are negligible, and
both levels grow exponentially with their respective SR
rates. The transition terms in eq. \eqref{eq:two_level_dynamics}
become important when the transition rate starts to compete with the
growth rate. The occupation number of the excited level is maximized
when
\begin{eqnarray}
  N_g = \Gamma^{\rm{sr}}_{e}/\Gamma_{t},
\end{eqnarray}
after which the excited state depopulates rapidly. The size of
the signal depends on whether $\Gamma^{\rm{sr}}_{e}>
\Gamma^{\rm{sr}}_{g}$ or vice-versa. An example of the $N_g, N_e$ and $h$
time evolution for the two cases is shown in
fig. \ref{fig:two-level-sys}.

If $\Gamma^{\rm{sr}}_{e}> \Gamma^{\rm{sr}}_{g}$
(fig.~\ref{fig:two-level-sys}, left), $N_{e}\gg N_{g}$ at the time when the transition
terms become relevant. The transition GW strain
keeps growing as the excited level gets depleted, until both levels are
populated with an equal number of axions. After that, the signal drops
precipitously as the excited level empties out and the ground state
returns to growing with rate $\Gamma^{\rm{sr}}_{g}$.

If $\Gamma^{\rm{sr}}_{g}> \Gamma^{\rm{sr}}_{e}$
(fig.~\ref{fig:two-level-sys}, right), the excited level depopulates very
quickly once it reaches the maximum, so the transition term for the
ground state is never important. The smaller occupation number of the excited level
suppresses the overall GW strain: the peak transition strain is
smaller compared to the previous case by an additional factor of
$\sqrt{
  {\Gamma^{\mathrm{sr}}_e}/{\Gamma_t}}^{|\Gamma^{\mathrm{sr}}_e-\Gamma^{\mathrm{sr}}_g|/\Gamma^{\mathrm{sr}}_g}\sim
\mathcal{O}(10^{-35})^{{|\Gamma^{\mathrm{sr}}_e-\Gamma^{\mathrm{sr}}_g|/\Gamma^{\mathrm{sr}}_g}}$.

In both cases, the transition process has a characteristic
timescale of $\Gamma_{e,g}^{\rm{sr}-1}$, typically decades for
stellar BHs. The maximal occupation numbers are
controlled by the ratio $\Gamma^{\mathrm{sr}}_{e}/\Gamma_t$, and
the peak strain is proportional to the superradiance rate:
\begin{equation}
  h_{\rm{peak}} \propto
  \frac{\Gamma^{\mathrm{sr}}_e}{\sqrt{\omega_{\rm{tr}}\Gamma_t}}.
\end{equation}
The maximum signal occurs before the occupation numbers reach
$N_{\rm{max}}$; the ground state will continue to grow
exponentially until all available angular momentum is extracted and
the SR condition is no longer satisfied. At this point,
annihilations become the dominant process. For general light bosons, the transition
signal strain is unaffected as long as $N^{\rm{peak}}_{e,g}<N_{\rm{bosenova}}$, or
$f_a\gtrsim 10^{14}\gev$.

We assume above that the initial number of axions
occupying each level is ${N_e}_0={N_g}_0=1$, as expected when
superradiance starts from scratch. If it restarts after being
disturbed e.g. by a bosenova which leaves one level partially
occupied, the other level does not have time to grow to the optimal
occupation number and the transition rate is not significant. This
implies that the transition signal will most likely appear only once in the
lifetime of a stellar black hole.

\begin{figure}[t]
  \centering
  \includegraphics[width=\linewidth]{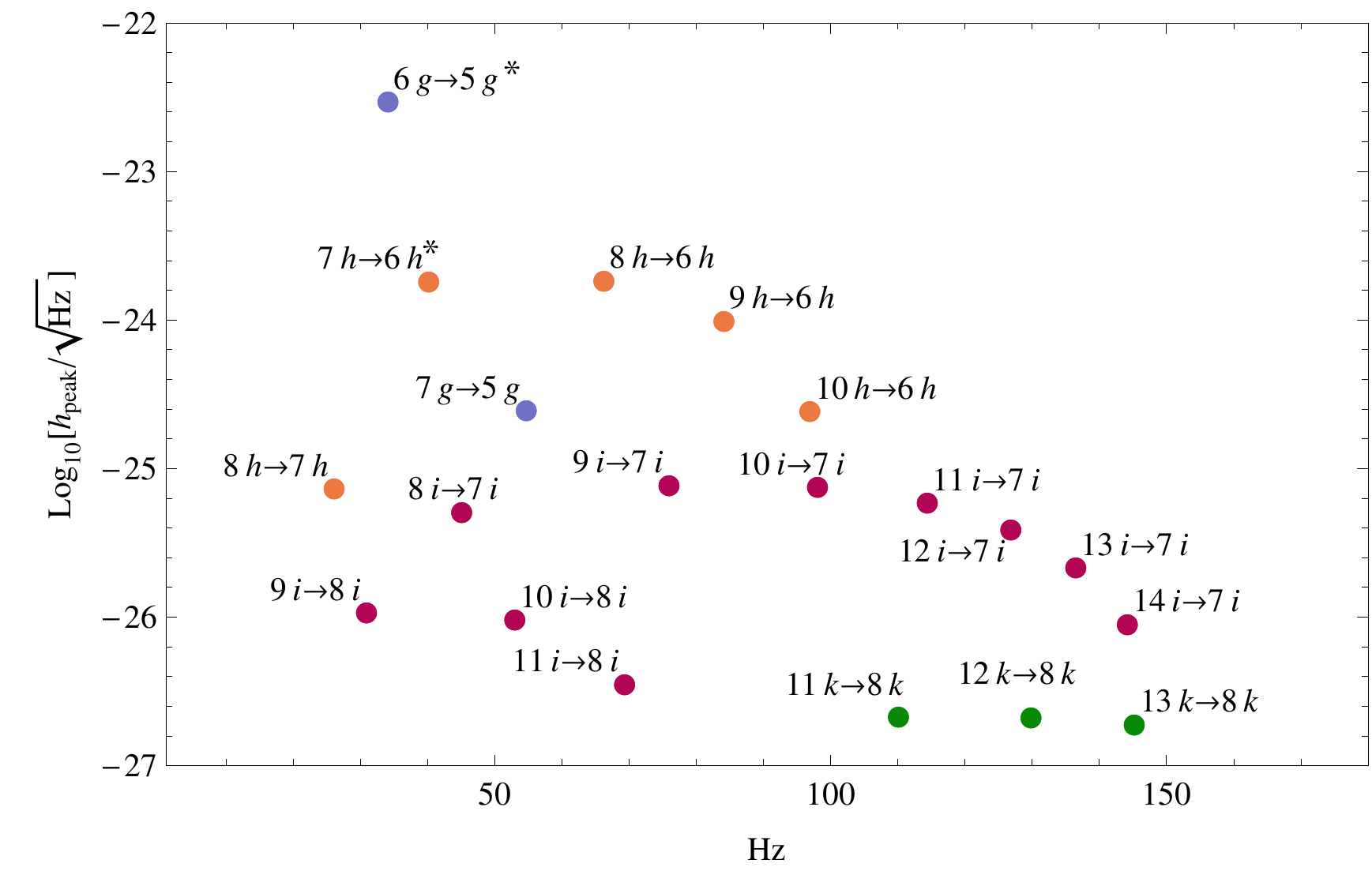}
  \caption{Transition signal strains for different level transitions from a $10 \msun$ black hole
    system 10 kpc away ($a_* = 0.99$, $\alpha/\ell = 0.3$), assuming
    $25$~hr integration time. The bottom axis shows the corresponding GW frequency. We focus on the starred ($^*$)
    transitions as most promising for GW detection. The strain shown
    here is approximate; we also make the simplistic assumption that
    in each case only two SR rates dominate. }
  \label{fig:translevels}
\end{figure}

Fig.~\ref{fig:translevels} shows the relative GW strains of various
transitions for a BH of mass $10 \msun$.  The analysis above shows
that transitions are relevant for the evolution of superradiance only
when the two levels can be simultaneously populated. The most
promising cases for transition signals are the ones with the smallest
difference between superradiance rates: $\Delta n \neq 0$ and $\Delta
\ell=\Delta m= 0$.  The $\ell=4$ (``$g$'') levels are ones with lowest
$\ell$ that satisfy $\Gamma^{\mathrm{sr}}_{e}>
\Gamma^{\mathrm{sr}}_{g}$ to avoid the suppression factor discussed
above; levels with higher $\ell$ have lower superradiance rates and
correspondingly lower peak strains. When three or more levels have
similar superradiance rates, the transitions between them may be
inhibited; such situations require further analysis. For instance, the
$7h\rightarrow 6h$ transition suppresses the $8h\rightarrow 6h$
transition power (suppression not shown in fig.~\ref{fig:translevels}).

We see that the $6g\rightarrow 5g$ transition is the most likely to be
seen by GW detectors, followed by $7h\rightarrow 6h$; these are the
levels that we use in our signal estimates below.

\subsubsection{Advanced LIGO/VIRGO Prospects}
\label{sec:transition_ligo}
Transitions between superradiant levels around stellar black holes
fall in the sensitivity band of Advanced LIGO and VIRGO: they have
frequency $f~\sim~15\,\mathrm{Hz}\times({\mu_a}/ 10^{-11}\ev)$ and
peak strains as high as $h~\sim~10^{-24}$ for a BH $10$ kpc away. A
search for such a signal with GW observatories is very promising,
especially since the GW emission is monochromatic. The length of the
signal for most of the observable parameter space is $10$ years or
more, so if a signal is detectable on Earth, it will persist 
longer than an observatory's science run.
\begin{figure}[t]
  \begin{center}
    \includegraphics[trim = 0mm 0mm 0mm 0mm, clip, width
    =1.02\linewidth]{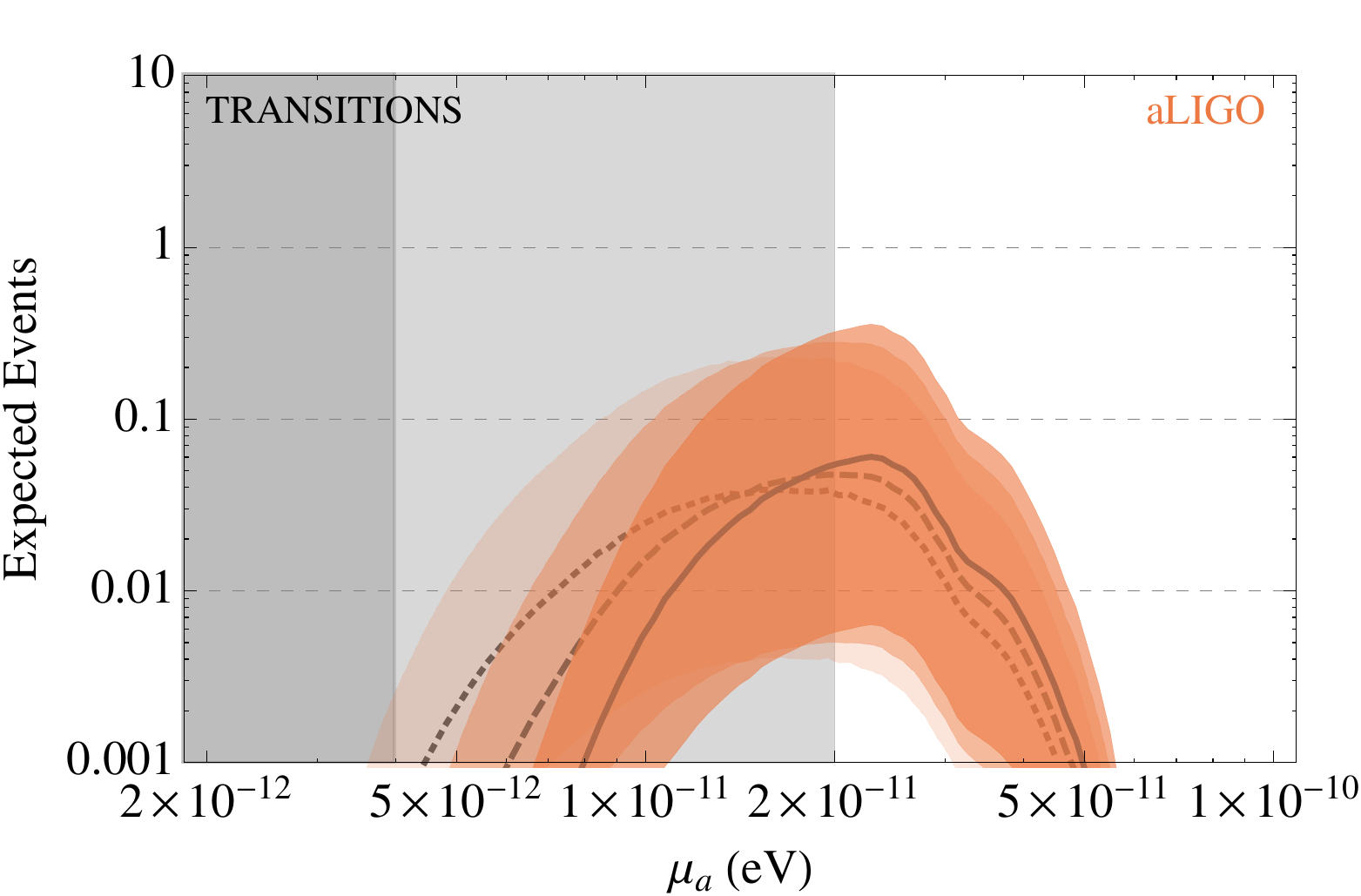}
    \vspace{-2.5mm}
    \caption{Number of $6g\rightarrow 5g$ and $7h\rightarrow 6h$
      transition events expected at aLIGO as a function of the axion
      mass, assuming a monochromatic search with $121\times 250$ hr
      integration time and $C_{\rm{tf}} =10$. Each event typically
      lasts several decades. We assume $4.1\msun$ minimum BH mass; if
      the minimum BH mass is smaller the curves would shift to higher
      axion masses.  The three lines correspond to varying the BH mass
      distribution width, from narrow (solid) to wide (dotted).  The
      bands around the central curves correspond to optimistic and
      pessimistic estimates of other astrophysical uncertainties (see
      text). The vertical shaded regions are disfavored by the
      observation of rapidly spinning BHs, for bosons with coupling
      equal to that of the QCD axion (light gray) or stronger (dark
      gray) (see section~\ref{sec:spin_limit}).}
    \label{fig:aLIGOreach}
    \vspace{-2.5mm}
  \end{center}
\end{figure}

The optimal search strategy is similar to an existing search for
periodic gravitational waves from e.g. asymmetric neutron stars
\cite{aasi2013einstein}.  We base our estimates on the design aLIGO
noise level \cite{Aasi:2013wya}; we expect similar reach for Advanced
VIRGO. The signal from a BH with $a_* = 0.99$ is visible up to
$10\mpc$ away, and one with $a_* = 0.9$ up to $30 \kpc$ away;
transition signal scales with the superradiance rate, so a black hole
with larger spin can be seen from further away. The best reach
is for masses around $3\times 10^{-11}\ev$ (see fig.~\ref{fig:treach}
in app.~\ref{sec:app_exp_reach}).

Given the promising reach, we can estimate the number of events aLIGO
could observe, shown in fig.~\ref{fig:aLIGOreach}.  To quantify the
event rate, we consider the probability to find a BH with high spin
and mass in the appropriate range to lead to transitions, as well as
the number and spatial distributions of BHs in our neighborhood young
enough to be undergoing transitions today.

\begin{figure}[t]
  \begin{center}
    \includegraphics[trim = 0mm 0mm 0mm 0mm, clip, width
    =1.02\linewidth]{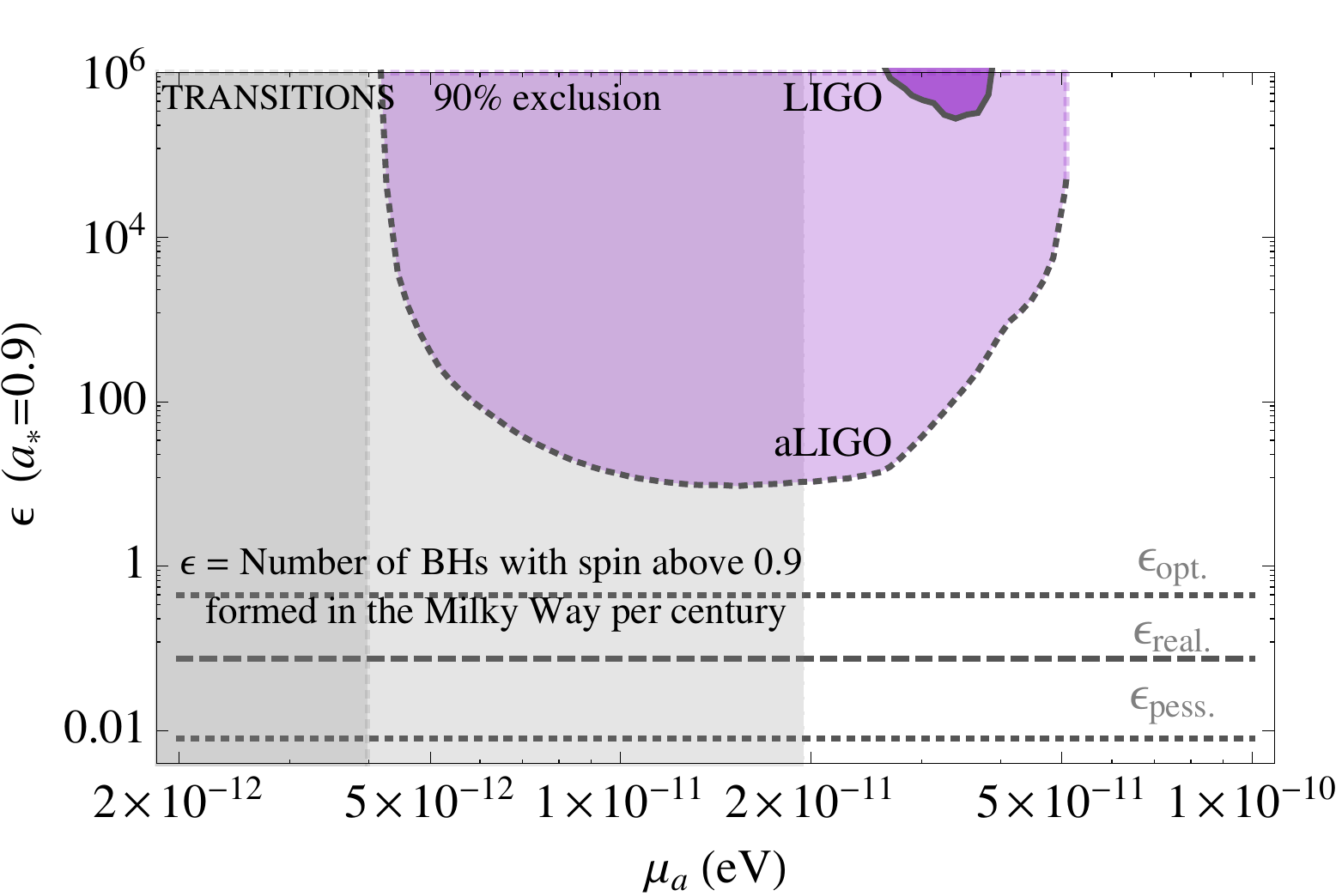}
    \vspace{-2.5mm}
    \caption{LIGO and projected aLIGO exclusion for $6g\rightarrow 5g$
      and $7h\rightarrow 6h$ transitions. The vertical shaded regions
      are the same as in fig.~\ref{fig:aLIGOreach}. We assume BH mass
      and distance distribution as described in the text. For the LIGO
      curve, we use the same integration time as the current LIGO
      monochromatic search, $N_{\mathrm{seg}}=121$, $T_{\mathrm{coh}}
      =25$ hrs, and $C_{\rm{tf}} =20$; for aLIGO, we use
      $N_{\mathrm{seg}}=121$, $T_{\mathrm{coh}} =250$ hrs, and
      $C_{\rm{tf}} =10$. The horizontal lines indicate optimistic,
      realistic, and pessimistic values for the $\epsilon$-parameter.}
    \label{fig:aLIGOtrexcl}
    \vspace{-2.5mm}
  \end{center}
\end{figure}

We use optimistic, realistic, and pessimistic estimates for the
astrophysical distributions following the event rate estimates of the
LIGO collaboration \cite{Abadie:2010cf}. We fold the following astrophysical
distributions of stellar BHs into our transition event rate estimates at
aLIGO (see app.~\ref{sec:app_bh_properties} for details):
\begin{itemize}
\item Mass distribution: we use a fit to data with a minimum BH mass
  and exponential drop toward high masses characterized by
  $M_0=4.7^{+3.2}_{-1.9} \msun$ \cite{Farr:2010tu}. In
  fig.~\ref{fig:aLIGOreach} we use a minimum mass of $ 4.1 \msun$; the
  events shift to higher axion masses if the minimum BH mass is
  smaller. We show the central and $1\sigma$ widths of the mass
  distribution in fig.~\ref{fig:aLIGOreach}.

\item {Spin distribution}: we take the measured distribution, $30\%$
  of $a_*>0.8$, as a realistic estimate for natal spins. We consider
  $90\%$ above $0.9$ optimistic and a flat distribution pessimistic.

\item {Formation rate}: barring rare violent events, the maximal
  transition signal occurs once per stellar BH lifetime, and the
  event rate is directly proportional to the BH formation rate. We estimate
  the BH birth rate to be $0.38^{+0.52}_{-0.3}$ per century based on
  supernova rates \cite{Diehl:2006cf,Prantzos:2003ph} and BH fraction
  of supernova remnants \cite{heger2003massive}.

\item {Distance distribution}: we assume the spatial distribution of
  BHs is proportional to the stellar distribution inside
  \cite{sale2010structure} and outside \cite{kopparapu2008host} the
  Milky Way. The event rates are dominated by black holes
  near the galactic center.
\end{itemize}

We can use the event rates to constrain a combination of the axion mass and
astrophysical parameters. We isolate the most relevant astrophysical
uncertainties, the BH formation rate and the spin distribution, and
define
\begin{equation}
  \label{eq:17}
  \epsilon \equiv \mbox{(BH formation rate)}\times \left(\,
    \begin{aligned}
      &\mbox{Fraction of BHs}\\
      &\mbox{with $a_* > 0.9$}
    \end{aligned}\,\right).
\end{equation}
The $90\%$ exclusion in the $\ma\,\rm{vs.}\,\epsilon$ plane is shown
in fig.~\ref{fig:aLIGOtrexcl}, fixing the BH mass and distance
distribution to the central values discussed above. LIGO is not yet
sensitive to reasonable astrophysical values for $\epsilon$, but aLIGO
will make considerable progress toward probing interesting parameter
space.

Unlike the case of neutron stars which spin
down due to GW emission, the transition signal's frequency is set by
the level splitting and is constant up to corrections from the
nonlinearities of the cloud itself. For the QCD axion, the maximum
signal occurs when the occupation number of the cloud is much smaller
than the nonlinear regime, resulting in a tiny frequency drift,
\begin{equation}
  \label{eq:3}
  \!\! \!\!\frac{df}{dt}\!\!\simeq 10^{-11}\!\frac{\hz}{\mathrm{s}}\!
  \left(\!\frac{f}{90\hz}\!\right)\!
  \!\left(\!\frac{M}{10 \msun}\!\right)\!\!
  \left(\!\frac{10^{17}\!\gev}{f_a}\!\right)^{\!2}\!\!
  \left(\!\frac{5\mathrm{yr}}{T}\!\right)^{\!2}\!\!
\end{equation}
where $T \gtrsim 5 \yr$ is the characteristic signal length, set by
$\Gamma_{\rm{sr}}^{-1}$; most of our parameter space has smaller
frequency drift, down to $10^{-20} \hz\!/\mathrm{s}$, and the signal
is well-approximated as having a constant frequency.

While the change in frequency is small, observing and correlating it
with the signal amplitude can provide additional handles on the
magnitude and sign of the particle's self-interaction. The amplitude
and sign of the frequency drift are correlated with the amplitude of
the signal. If $\Gamma^{\rm{sr}}_{g}> \Gamma^{\rm{sr}}_{e}$, the
frequency of the emitted graviton increases with time.  If
$\Gamma^{\rm{sr}}_{g} < \Gamma^{\rm{sr}}_{e}$, the frequency decreases
as both levels grow and then increases as the excited level
empties. For particles with repulsive interactions, the drift is in
the opposite direction.

The theoretical uncertainties of the expected event rates and
exclusions depend on the SR rate and the transition rate, the precise
values of which require further numerical study to include
higher order effects of the metric and deviations from the hydrogen
wavefunctions which we assume in our estimates. The dependence on the SR rate is
mild since a larger SR rate leads to larger signal strain but shorter
signal length. The dependence on the transition rate is stronger;
event rates and exclusions scale with $\sim \Gamma_t^{-1/2}$, which we
expect to have uncertainties of $\mathcal{O}(1)$.

If there is an axion or light boson close to $10^{-11}\ev$ in mass,
with some luck, aLIGO could see a monochromatic signal lasting for many years. 

\subsubsection{Future Gravitational Wave Observatories}
\label{sec:futuretransitions}

Upcoming observatories such as Advanced LIGO and VIRGO are perfectly
suited to search for superradiance signals from stellar black holes;
to detect SR signals from supermassive black holes (SMBHs), we look to
future, lower-frequency proposals: eLISA and
AGIS. The eLISA observatory is a laser interferometry gravitation wave detector~\cite{LISA}; LISA
Pathfinder is planned to test mile-stone requirements in its
experimental development~\cite{pathfinder}. AGIS is a recently
proposed single-baseline gravitational wave detector based on atom
interferometry~\cite{Dimopoulos:2008sv}. This promising new idea is
currently under development and could exceed light-based
interferometer sensitivities \cite{2015arXiv150106797H,Graham:2012sy}.

The reach of eLISA and AGIS is promising, extending to as far as a 
hundred Mpc, but the best detector sensitivity falls in the range of
intermediate-mass BHs, $M\lesssim 10^5\msun$ (see
fig.~\ref{fig:treach_sm} in app.~\ref{sec:app_exp_reach}). Lack of
estimates for distributions of intermediate mass BHs makes even an
approximate estimate of event rates difficult. In the limit where all
supermassive black holes have mass $M=10^6\msun$ and most have maximal
spin (see app.~\ref{sec:app_bh_properties}), low-frequency detectors
can observe up to $100$ BHs undergoing transitions.

\subsection{Annihilations}
\label{sec:annihilation}

Another source of gravitational wave emission is axions
annihilating to gravitons in the black hole background,
$a+a \rightarrow g+g_{\rm{bg}}$; this process is analogous to
electron-positron annihilation to a photon in the background of a
nucleus \cite{PhysRev.44.510.2}. The GW frequency is
\begin{equation}
  \label{eq:15297}
  \omega_{\rm{ann}} = 2 \omega_{a} \simeq 2 \ma \left(1-\frac{\alpha^2}{2
      n^2}\right). 
\end{equation}

When a single level dominates the evolution of the axion-BH system,
its occupation number $N(t)$ grows with its SR rate while being
depleted by axions pair annihilating into gravitons,
\begin{equation}
  \label{eq:annih_dynamics}
  \frac{d N}{d t} = \Gamma_{\rm{sr}}N - \Gamma_a N^2.
\end{equation}
Here, $\Gamma_a$ is the annihilation rate for one pair of axions, of order
\begin{equation}
  \label{eq:annih_rate}
  \Gamma_{a} \simeq 10^{-10}\left[\left(\frac{\alpha/\ell}{0.5}\right)^{p} +
    \mathcal{O}\left(\frac{\alpha/\ell}{0.5}\right)^{p+1}\right]\frac{G_{N}}{r_g^3},
\end{equation}
where $p = 17$ for $\ell = 1$ and $p = 4\ell+11$ for $\ell \geq 2$;
see app.~\ref{sec:app_GW_power} for full expressions and a discussion
of the $\alpha$ dependence of the $\ell=1$ level. We only consider $n
= \ell + 1$ in this section. The annihilation rates close to the
superradiance boundary ($\alpha/\ell = 1/2$) are similar for all
$\ell$-levels. At smaller $\alpha/\ell$, the annihilation rate is
velocity suppressed, with the suppression more pronounced at higher
$\ell$.

Comparing eqs.~\eqref{eq:transition_rate} and \eqref{eq:annih_rate}
we see that annihilation is the slowest process, $\Gamma_a\ll
\Gamma_{t}\lll \Gamma_{\mathrm{sr}}$. Nevertheless,
annihilations are important when the occupation number of a single
level is far larger than that of the others, as is the case when a single
superradiance rate dominates.

When $N < \Gamma_{\rm{sr}}/\Gamma_a$, the axion population grows
exponentially with the superradiance rate. Once the axion cloud
extracts the maximum possible spin from the BH, $N(t)=N_{\rm{max}}$,
superradiance shuts off and the occupation number evolves as:
\begin{equation}
  N(t) = \frac{N_{\rm{max}}}{1+\Gamma_a N_{\rm{max}}t}.
  \label{eq:annihtime}
\end{equation}
The corresponding gravitational wave signal strain is
\begin{equation}
  h(t)   = N(t)\sqrt{\frac{4 G_N}{r^2\omega_{\rm{ann}}}\Gamma_a}.
\end{equation}
Both the peak strain,
\begin{align}
  \label{eq:2}
  h_{\rm{peak}}
  &\simeq 10^{-22}\left(\frac{1 \kpc}{r}\right) \left(\frac{\alpha/\ell}{0.5}\right)^{\frac{p}{2}} \frac{\alpha^{-\frac{1}{2}}}{\ell}\left(\frac{M}{10\msun}\right),
\end{align}
and peak duration, $ (N_{\rm{max}}\Gamma_a)^{-1}$, are independent of
the superradiance rate. For stellar BHs, the signal at its peak value can last for
thousands of years.
\begin{figure}[t]
  \begin{center}
    \includegraphics[trim = 0mm 0mm 0mm 0mm, clip, width
    =\linewidth]{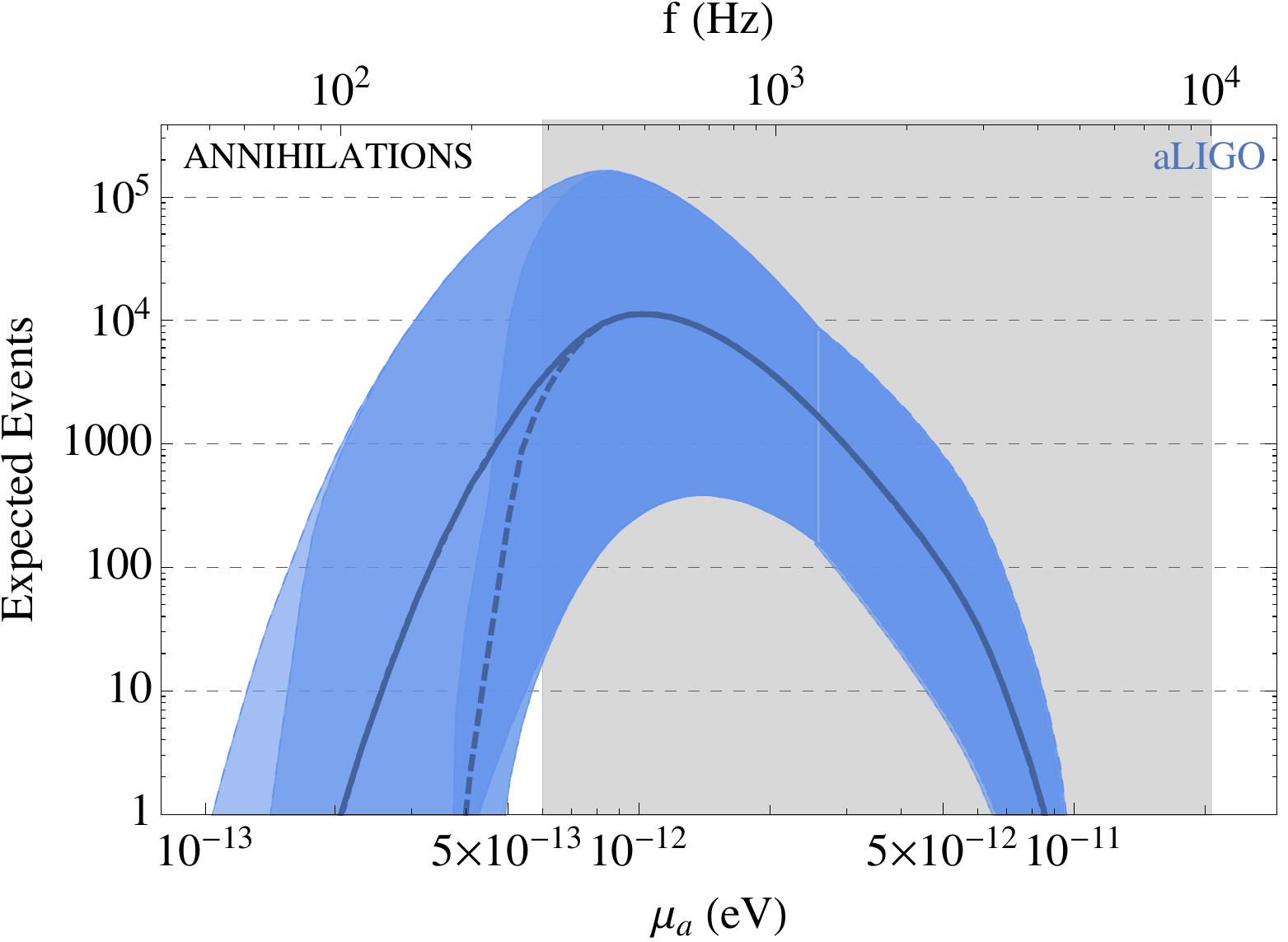}
    \vspace{-2.5mm}
    \caption{Number of $2p$ annihilation events possible to observe
      with aLIGO as a function of the axion mass, assuming a
      monochromatic search with $121\times 250$ hr integration time
      and $C_{\rm{tf}} =10$. Each event lasts thousands of years or
      longer. The vertical shaded region is disfavored by black hole
      spin measurements assuming QCD axion coupling strength. Each of
      the three bands corresponds to cutting off the BH mass
      distribution at a maximum mass of $\{30,80,160\} \msun$ (dark,
      medium, and light blue) including optimistic and pessimistic
      estimates of astrophysical uncertainties, with central values
      given by the dashed ($M_{\rm{max}} =30 \msun$) and solid ($M_{\rm{max}}
      =80,\,160 \msun$) curves (see text). } 
    \label{fig:aLIGOreach_ann}
    \vspace{-2.5mm}
  \end{center}
\end{figure}

For supermassive black holes, we expect the BH to move adiabatically
on the Regge trajectory since $(N_{\rm{max}}\Gamma_a)^{-1}$ is of
order $10^9$ years, comparable to the Eddington accretion time. The
motion of the BH along the Regge trajectory may be interrupted by
in-falling stellar mass BHs or neutron stars; we estimate that such
events occur every $10^6-10^7$ years based on the star infall rate in
\cite{Khabibullin:2014iaa} and that about one in 100 stars is a BH or
neutron star \cite{brown1994scenario}. When this happens,
$\mathcal{O}(1)$ of the cloud falls back into the black hole,
restoring the spin almost back to the value that it would have had
without superradiance. We require that the superradiance rate is much
larger than the violent infall event rate; then except for short
intervals around the time of the infall, the signal coming from
annihilations will be close to maximal until the black hole grows in
mass such that it is no longer affected by superradiance. While the
infall rate is uncertain, it only affects a small part of the
parameter space. The size of the cloud is determined by the difference
between the spin the BH would have had without superradiance and the
spin corresponding to its mass on the trajectory.
Fig.~\ref{fig:lisaevents} takes into account these subtleties for
supermassive black holes.

So far, we have assumed that the axion's self-interaction is of QCD axion
strength. If the interaction is stronger such that bosenovae are
relevant, the axion cloud only grows to $ N_{\rm{bosenova}}<
N_{\rm{max}}$, and the maximum annihilation signal is proportionally
smaller and lasts for a shorter time.

\subsubsection{Advanced LIGO/VIRGO Prospects}

\begin{figure}[t]
  \begin{center}
    \includegraphics[trim = 0mm 0mm 0mm 0mm, clip, width
    =1.02\linewidth]{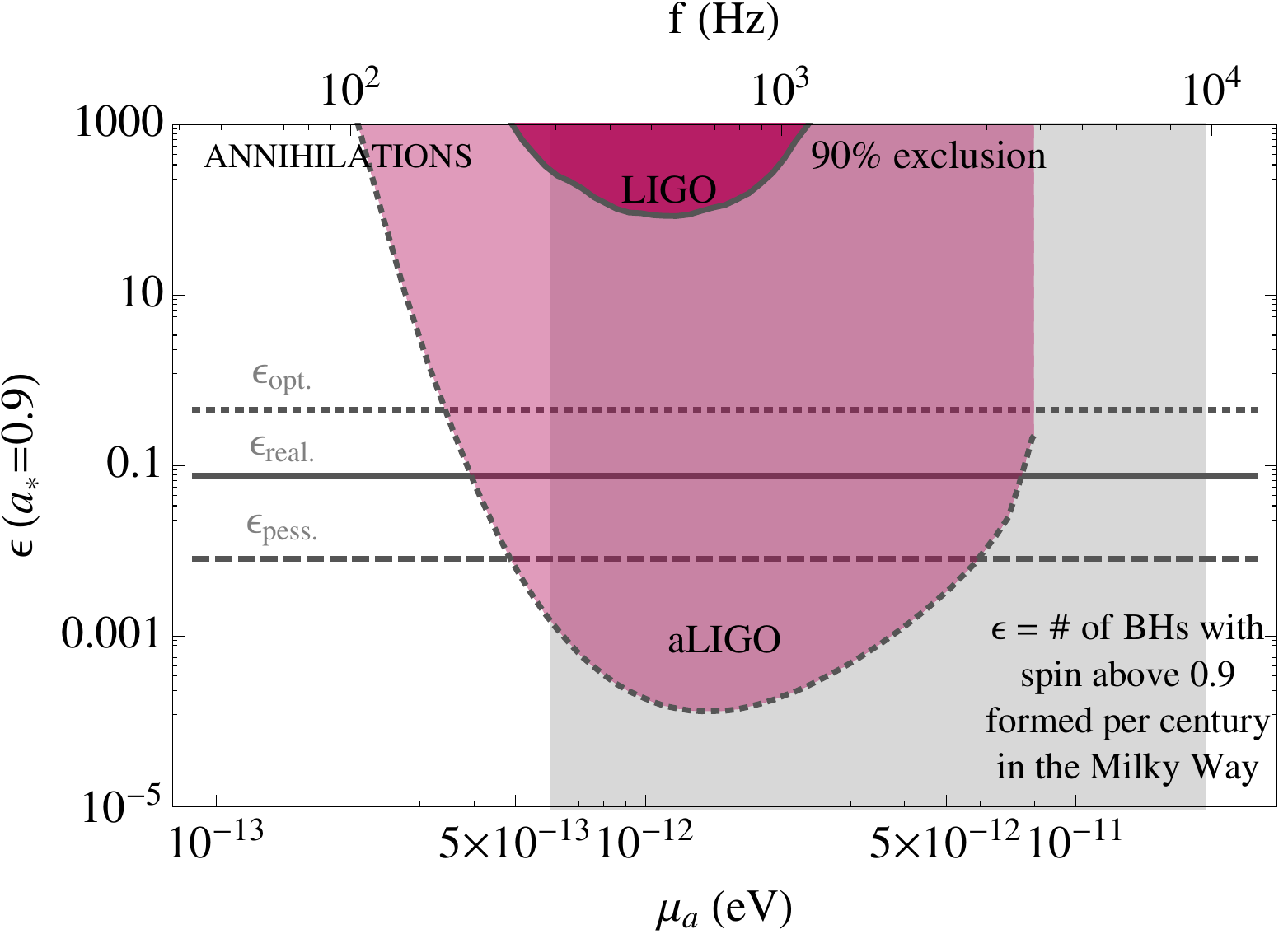}
    \vspace{-3mm}
    \caption{LIGO and projected aLIGO exclusion for $2p$
      annihilations. The vertical shaded region is the same as in
      fig.~\ref{fig:aLIGOreach_ann}. We assume BH mass and distance
      distribution as described in text. For the LIGO curve, we use
      the same integration time as the current LIGO monochromatic
      search, $N_{\mathrm{seg}}=121$, $T_{\mathrm{coh}} =25$ hrs, and
      $C_{\rm{tf}} =20$; for aLIGO, we use $N_{\mathrm{seg}}=121$,
      $T_{\mathrm{coh}} =250$ hrs, and $C_{\rm{tf}} =10$.  The
      horizontal lines indicate optimistic, realistic, and pessimistic
      values for the $\epsilon$-parameter. }
    \label{fig:aLIGOannexcl}
    \vspace{-3mm}
  \end{center}
\end{figure}

The annihilation signal is quite distinctive: it is monochromatic with
frequency of twice the axion mass, $f~=~10\,\mathrm{kHz}\times
({\mu_a}/{10^{-11}\ev})$, possibly lasting for thousands of
years. The optimal reach of aLIGO is for axion masses around $5\times 10^{-13}\ev$
which correspond to intermediate mass black holes, $M>30\msun$ (see
fig.~\ref{fig:areach} in app.~\ref{sec:app_exp_reach}); these are poorly understood and have
only been recently observed \cite{pasham2014400}.

To estimate the event rate, we use the mass distribution of stellar
BHs, which includes an exponential tail extending to heavier BHs.  In
fig.~\ref{fig:aLIGOreach_ann}, we estimate the event rate expected in
aLIGO from axion annihilations using the same BH astrophysical
distributions as in section~\ref{sec:transition_ligo}. We folded the
BH mass distribution width into the optimistic and pessimistic
estimates, and we indicate the expected event rate for
different maximum stellar BH mass cutoffs (see
app. \ref{sec:app_bh_properties}). Especially at light axion masses,
the signal is subject to large astrophysical uncertainties of heavier
($30$--$100\msun$) BH mass distributions, and the difference between the
optimistic and pessimistic estimate is dominated by the shape of the
exponential tail of the BH mass distribution.

The event rates range from $\mathcal{O}(1)$ to $\mathcal{O}(10^4)$;
part of the parameter space with appreciable event rates is disfavored
by BH spin measurements (section \ref{sec:spin_limit}). The event
rates for axion masses lighter than the excluded range are very
promising, with possibly thousands of monochromatic signals due to a
Planck-scale QCD axion or another boson in the same mass range.

Similarly to our exclusion in fig.~\ref{fig:aLIGOtrexcl} for
transition signals, we place an exclusion in the $\epsilon$
(eq.~\eqref{eq:17}) vs. axion mass plane in
fig.~\ref{fig:aLIGOannexcl}, fixing the distance distribution as
discussed above and using a conservative mass distribution with a
narrow width and upper BH mass cutoff of $80\msun$.  The reach for
annihilations covers much of the region disfavored by black hole spin
measurements and can provide a cross-check for the exclusion. The
large event rates make detection in the mass range of $1-6\times
10^{-13}\ev$ possible, and aLIGO can probe a meaningful region in the
axion parameter space.

Advanced LIGO is sensitive largely to signals from within the Milky
Way. An increase in detector sensitivity by a factor of $10$ (such as
the Einstein Telescope \cite{aasi2013einstein}) would increase the
number of events by a factor of $\mathcal{O}(10)$: the detector would
be sensitive to signals with smaller strain which last for a
proportionally longer time. To reach a cubic scaling with distance the
detector reach must be $ >30 \mpc$, at which point the density of
galaxies scales in proportion to the volume. This would require a
detector with sensitivity a factor of $100$ better than aLIGO.

The theoretical uncertainties in the expected annihilation event rates
are independent of the superradiance rate, while an $\mathcal{O}(10)$
increase in the annihilation rate extends the reach to lower $\ma$ by
$\sim 20\%$. The change is not significant because a higher signal
strain is compensated by a shorter signal length. As explained in
app. \ref{sec:app_GW_power}, we expect the $2p$ annihilation rate to
have a weaker $\alpha$-dependence than our conservative analytical
estimate. Changing the $\alpha$-dependence would extend the event and
exclusion curves by a factor of $\sim 2$ toward lighter $\ma$ and
increase the peak event rates by a factor of $\sim 4$.

Large frequency drift can make a monochromatic search difficult. In
the case of annihilations, the signal grows to a maximum with the
superradiance rate, and the cloud then slowly depletes, resulting in a
small positive drift in frequency due to attractive self-interactions,
\begin{equation}
  \label{eq:15}
  \frac{df}{dt} \simeq 10^{-12}\frac{\hz}{\mathrm{s}}\!
  \left(\frac{f}{\kHz}\right)
  \left(\frac{M_{\rm{\small{Pl}}}}{f_a}\right)^2\!
  \left(\frac{10^3\, \mathrm{yr}}{T}\right),
\end{equation}
where $T = (\Gamma_a N_{\rm{max}})^{-1}$ is the typical length of the
signal. For most of the parameter space, the frequency drift is
smaller than $10^{-12}\hz\!/\mathrm{s}$. In our exclusion estimate
above, we only include signals that have frequency drift smaller than
$7\times 10^{-11} \hz\!/\mathrm{s}$ ($7\times 10^{{-}13}
\hz\!/\mathrm{s}$) for coherent integration time of $25$ ($250$)
hours based on the frequency binning in \cite{aasi2013einstein}. If
the experimental search techniques can accommodate higher frequency
drift, aLIGO can be sensitive to $\mathcal{O}(10\%)$ higher axion
masses.

If a frequency drift is detected, it can be used to
distinguish the signal from astrophysical sources. With the large
number of events expected, an additional tool to distinguish signal
from astrophysical sources is the detection of multiple monochromatic lines with
frequencies $\sim 2 \ma$, differing by corrections from
the $\mathcal{O}(\alpha^2)$ binding energy to the BH (eq.~\eqref{eq:15297}). 

The astrophysical uncertainties are large, but the event rates are
very promising. If a signal is identified, further study of its
amplitude and frequency as a function of time, correlated with
astrophysical measurements of the black hole source, could determine
the mass and coupling of the superradiating particle.

\subsubsection{Future Gravitational Wave Observatories}
\label{sec:annihilation_future}
\begin{figure}[t]
  \begin{center}
    \includegraphics[trim = 0mm 0mm 0mm 0mm, clip, width
    =.95\linewidth]{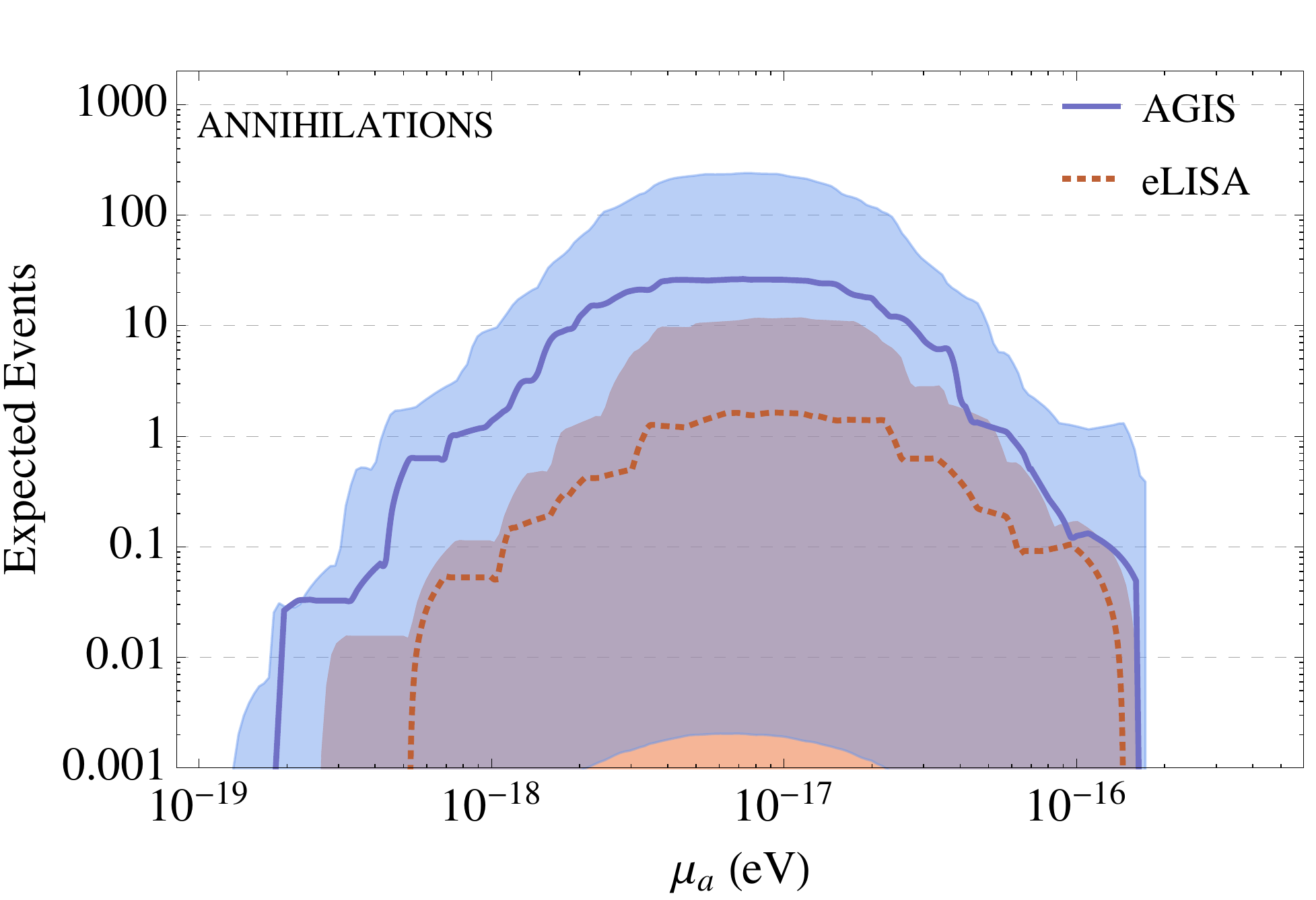}
    \vspace{-3mm}
    \caption{Expected number of events in eLISA (dashed) and AGIS
      (solid) as a function of the axion mass for SMBHs with axions in
      the $2p$-level undergoing annihilations.  Each event can last
      millions of years. The shaded bands bracket the optimistic and
      pessimistic estimates, dominated by SMBH spin distributions (see text).} \label{fig:lisaevents}
    \vspace{-5mm}
  \end{center}
\end{figure}
The frequency sensitivities of AGIS and eLISA are ideal for detection
of axion annihilations around supermassive black holes, with a
promising reach of up to $300 \mpc$ in distance for axions
between $10^{-18}-10^{-17}$ eV (see fig.~\ref{fig:areach_sm} in
app.\ref{sec:app_exp_reach}).

The masses and spins of supermassive black holes are determined by long
periods of accretion and so possible to estimate theoretically. We use
the following distributions for the event rates (more details in
app.~\ref{sec:app_bh_properties}):

\begin{itemize}
\item {Mass and distance distribution}: we use the distributions of
  \cite{kelly2012mass,shankar2009self}, with most SMBHs between
  $10^{6}$ and $10^{7}\msun$ in mass and about one $10^7\msun$ BH per
  Milky-Way type galaxy.

\item {Spin distribution}: The biggest uncertainty for event rates is
  due to the unknown spin distribution of SMBHs. We use a range of
  model predictions: optimistic, 70\% of SMBHs have spins
  $a_* \geq 0.93$ \cite{volonteri2005distribution}; intermediate, 70\%
  have spins $a_* \geq 0.7$ with 50\% above $ 0.9$ \cite{Sesana_2014};
  and pessimistic or low spin $a_* =0.2\pm 0.2$ \cite{King_2008}. The
  event rates are dominated by SMBHs with spins above
  $0.9$.\footnote{We thank referee B of PRD for proposing the above
    benchmarks for spin distributions and pointing us to the relevant
    references. See app.~\ref{sec:app_bh_properties} for further
    details.}
 
\item {Signal length}: most of the evolution of a SMBH which satisfies
  the SR condition is at the regime where accretion, superradiance,
  and axion annihilation happen adiabatically. This agrees with
  numerical results in \cite{Brito:2014wla}. In this case BH spin and
  axion cloud size remain in a steady-state with annihilations at the
  maximum rate.

\end{itemize}

We estimate the event rate expected by AGIS and eLISA in
fig.~\ref{fig:lisaevents}, giving the realistic as well as the
optimistic and pessimistic estimates based on the astrophysical
uncertainties above. The differences in sensitivity of the two
detectors are of order the astrophysical uncertainties.  If
there is a light boson with mass of  $10^{-18}-10^{-16}$
eV, the annihilation signal is dramatic, with thousands of continuous
events visible.

\subsection{Bosenovae}
\label{sec:bosenova}

A very different signature is the periodic collapse of the axion
cloud, known as a``bosenova'' in analogous condensed matter
systems. If the axion self-interaction is sufficiently strong ($f_a\ll
M_{\mathrm{pl}}$), the axion cloud will collapse at the critical size
$N_{\rm{bosenova}}$ before extracting all the BH's spin as allowed by the
superradiance condition. During the bosenova, a fraction of the cloud
falls into the black hole and the rest escapes to infinity, emitting a
gravitational wave burst.

The collapse lasts for approximately an infall time, $ t_{\rm{bn}} =
(c_{\mathrm{bn}} r_c)$, and has primary frequency component
\begin{equation}
  f_{\mathrm{bn}} \sim 30 \hz
  \left(\frac{16}{c_{\mathrm{bn}} }\right)\left(\frac{\alpha/\ell}{0.4}\right)^2\left(\frac{10 M_{\odot}}{M}\right),
\end{equation}
where $c_{\mathrm{bn}}$ parametrizes the collapse time 
($c_{\mathrm{bn}} \sim 16$ for the $2p$ level \cite{Yoshino:2012kn}).

Once the size of the cloud is reduced such that the system is no
longer nonlinear, the level grows again with its superradiance rate
until the next bosenova, and the growth-collapse cycle repeats until
the superradiance condition is no longer satisfied. The separation
between bursts depends on the fraction of the cloud which remains
bound to the BH after the bosenova event.

For example, the collapse of a typical axion cloud around a $10
M_{\odot}$ black hole with spin $a_* = 0.99$ will emit a burst lasting
$10^{-3}$~s.  If the axion coupling is e.g. $\,f_a= 6 \times
10^{16}\gev$, and each bosenova depletes the cloud to a size of
$10^{-4}N_{\rm{bosenova}}$, the signal will be in the distinct pattern
of $10$ spikes separated by quiet periods of $300$~s.

The strain of these periodic bursts can be large enough to be observed
by aLIGO: at a distance of a kpc, the quadrupole estimate gives a signal
strain of
\begin{equation}
  h\simeq\!10^{-21}\!
  \left(\frac{\sqrt{\epsilon}/c_{\rm{bn}}}{10^{-2}}\right)^2\!\left(\frac{\alpha/\ell}{0.4}\right)\left(\frac{M}{10
      M_{\odot}}\right) \left(\frac{f_a}{f_a^{\mathrm{max}}}\right)^2
\end{equation}
where $\epsilon$ is the fraction of the cloud falling into the BH
(\cite{Yoshino:2012kn} gives $\epsilon \sim 5\%$) and
$f_a^{\mathrm{max}}$ is the largest $f_a$ for which bosenovae 
occur, eq.~\eqref{eq:fa_bosenova}. For a $10\msun$ BH,
$f_a^{\mathrm{max}}$ corresponds to a small quartic coupling of
$\lambda \sim 10^{-77}$ for a generic scalar boson.

As we saw earlier, the QCD axion coupling is most likely too weak to
cause bosenovae around astrophysical black holes; light bosons with
larger self-couplings can lead to bosenovae that occur with a
tell-tale regularity and with signal frequency and strain that fall
promisingly in the range of aLIGO. Calculation of detector reach and
event rates are beyond the scope of this work; dedicated numerical
study is necessary to determine the precise shape, timing, and
amplitude of the bosenovae emission.

\section{Bounds from Black Hole Spin Measurements}
\label{sec:spin_limit}

Several techniques for spin measurements of black holes have been
developed and are constantly improving. The leading methods are
continuum fitting and X-ray relativistic reflection (for recent
reviews, see \cite{McClintock:2013vwa} and
\cite{Reynolds:2013qqa}). Both are based on the measurement of the
innermost stable circular orbit of the accretion disk: the radius
($R_{\rm{ISCO}}$) at which matter in the disk stops smoothly orbiting
and rapidly falls into the black hole. The $R_{\rm{ISCO}}$ is a
monotonically decreasing function of $a_*$ that becomes steeper for
$a_* \sim 1$, reducing error on high-spin measurements
\cite{bardeen1972rotating}. The systematics for converting
$R_{\rm{ISCO}}$ to a spin measurement are the same for both methods,
but the systematics for measuring $R_{\rm{ISCO}}$ are distinct.

Continuum fitting measures the $R_{\rm{ISCO}}$ through the temperature
and luminosity of the accretion disc. As the innermost stable circular
orbit gets closer to the BH horizon, the matter becomes denser and
hotter, increasing the luminosity of emitted radiation. The luminosity
does not depend in detail on the disk model: as matter orbits toward
the BH, its gravitational potential energy is converted to orbital
kinetic energy, with the amount radiated away determined only by
assumptions of steady state rotation, axisymmetry, and conservation
laws \cite{McClintock:2013vwa}. Just like the measurement of a star's radius
from its temperature and luminosity, the $R_{\rm{ISCO}}$ measurement
relies on the absolute distance to the BH and the disk inclination
with respect to line-of-sight. These, along with the BH mass needed to
convert $R_{\rm{ISCO}}$ to the dimensionless quantity $a_*$, lead to
the dominant sources of error in the spin measurement
\cite{McClintock:2013vwa,McClintock:2009dn,Miller:2009cw}. Sub-leading
uncertainties from disk modeling result in less than $10\%$ error in
$R_{\rm{ISCO}}$ and $<3\%$ errors in $a_*$ at high spins.  The
limitation is that the peak emission must be clearly visible,
excluding SMBHs for which emission is in the unobserved far-UV or soft
X-ray frequencies.

The X-ray relativistic reflection method (also known as Fe-K or broad
iron line method) measures the properties of the Fe-K$\alpha$ spectral
line, which is excited in the accretion disk by an external X-ray
source (e.g. the disk corona or the base of a jet); it can be used to
measure spins of both supermassive and stellar BHs. The X-rays are
partially absorbed, leading to an emission line from de-excitation
with a particularly distinctive shape. Little emission occurs inside
$R_{\rm{ISCO}}$ since there the density of matter drops sharply.  For
higher spin, matter can be closer to the horizon, resulting in an iron
line with a longer gravitationally red-shifted tail
\cite{fabian1989x}. If the disk is tilted, the rotation of the disk
Doppler-shifts the line to the blue and the red, resulting in two
peaks. Since iron has high abundance and
high fluorescence, and is isolated in the spectrum, the broadened and
distorted line can be fitted to find $R_{\rm{ISCO}}/r_g$ and the
inclination of the disk \cite{Reynolds:2013rva}. The Doppler and
gravitational shifts both depend on the dimensionless quantity
$R_{\rm{ISCO}}/r_g$, so knowledge of the BH mass is not needed to find
$a_*$.

\begin{figure}[t!]
  \centering
  \includegraphics[width=\linewidth]{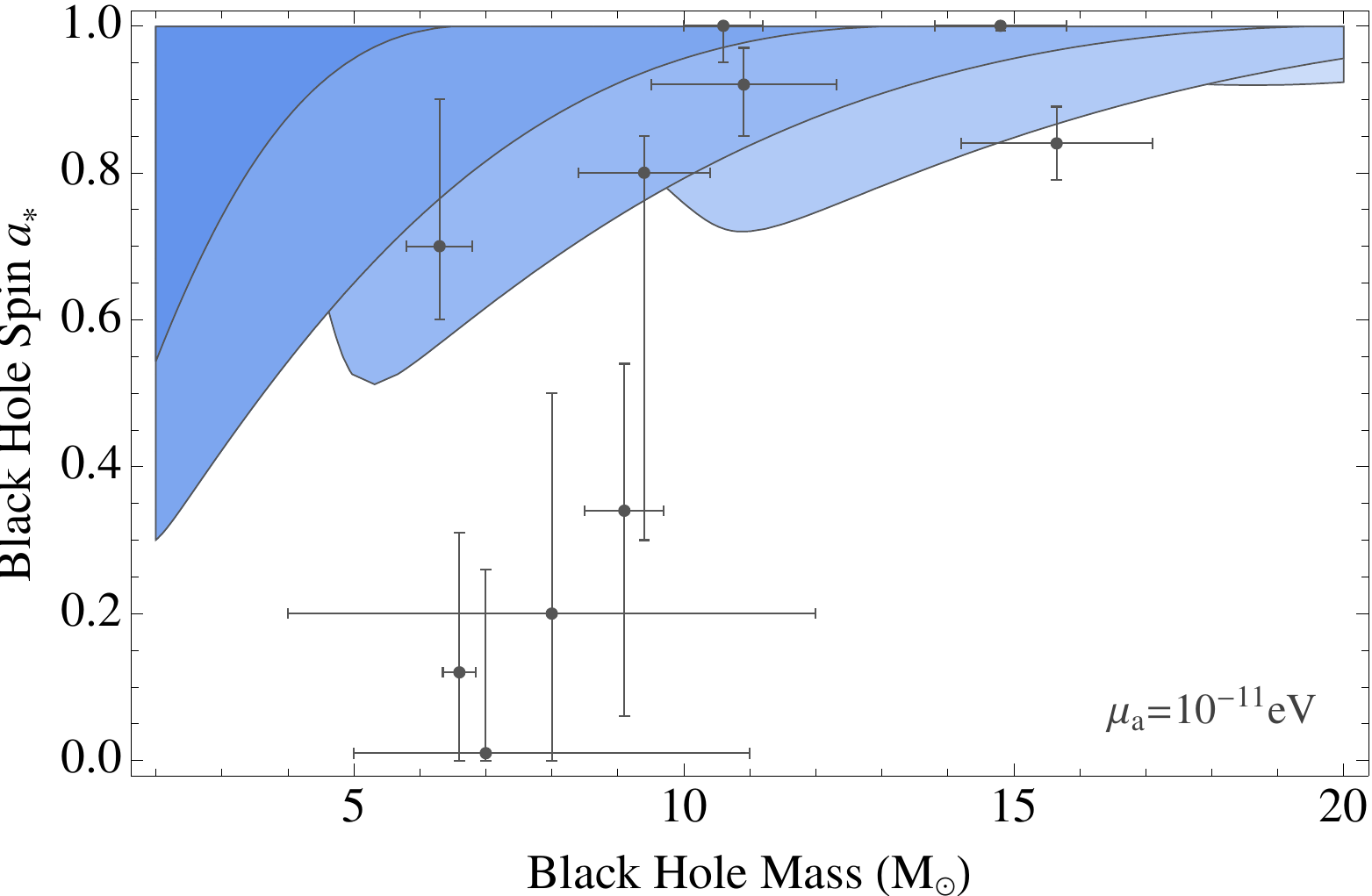}
  \caption{The shaded regions are affected by superradiance and
    bounded by Regge trajectories in the presence of a QCD axion with
    mass $\mu_a=10^{-11}\ev$. The points are stellar black holes
    measurements with $1\,\sigma$ error bars (for the two fastest
    spinning BHs, the $2\sigma$ lower spin bounds are shown).}
  \label{fig:regge_ex}
\end{figure} 

\SkipTocEntry\subsection*{Spin Limits}

\begin{figure}[t]
  \centering
  \vspace{-.4cm}
  \subfigure[ ]{
    \includegraphics[width=\linewidth]{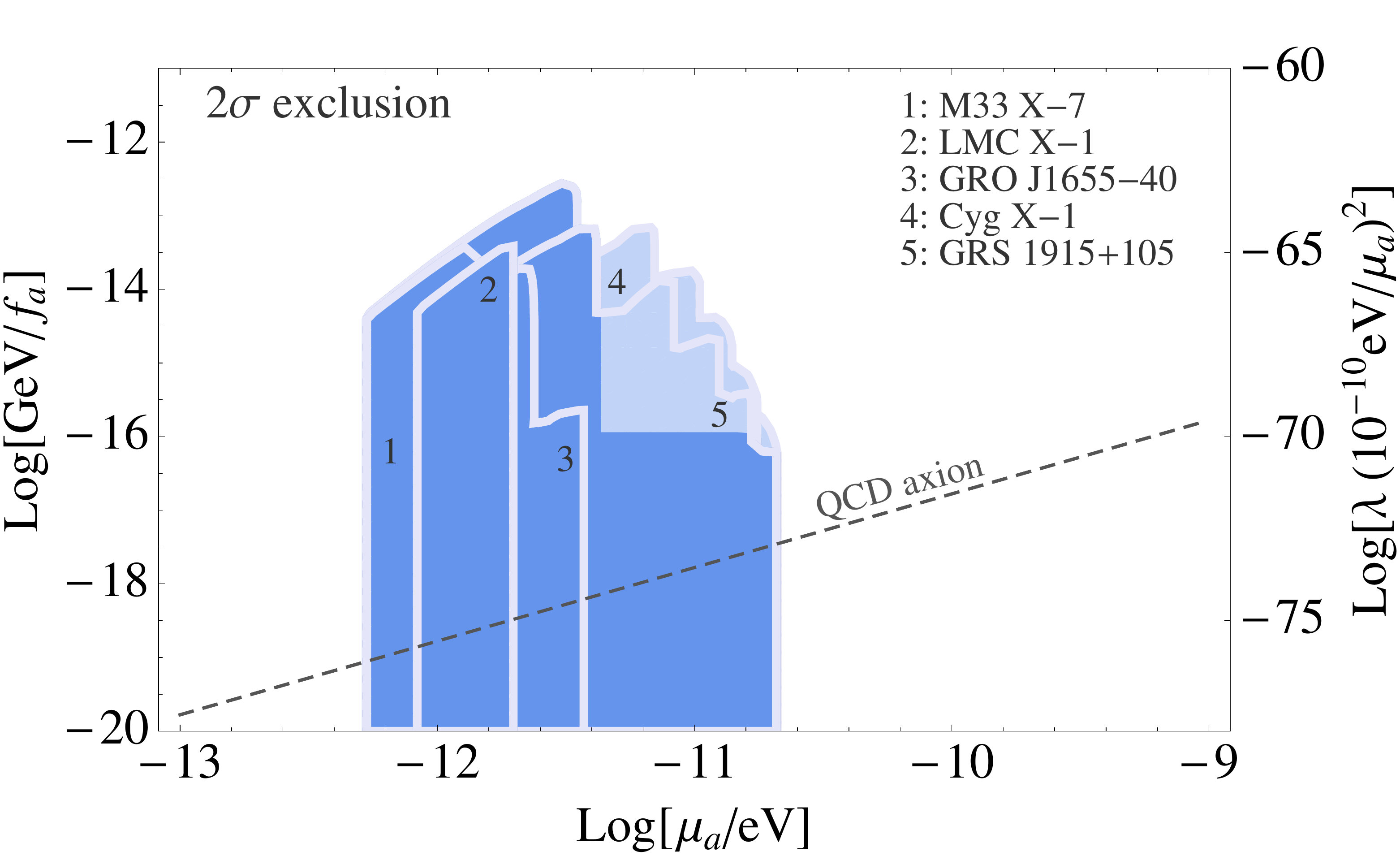}
    \label{fig:spinlim}
  }
  \subfigure[]{
    \includegraphics[width=\linewidth]{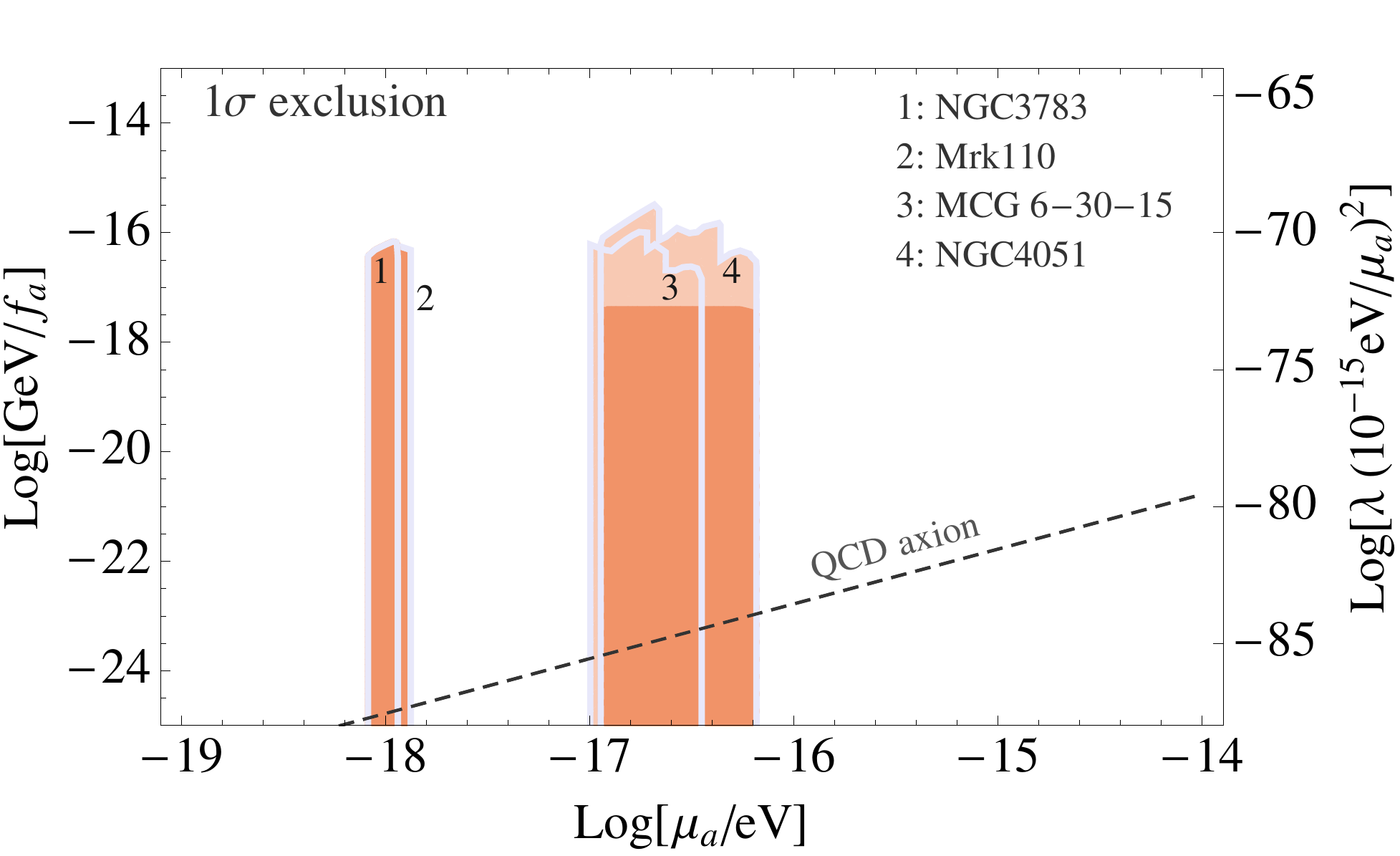}
    \label{fig:SMspinlim}
  }
  \caption{Limits on mass and self-coupling of light axions derived
    from (a) quickly rotating stellar black holes (at $2\sigma$) and
    (b) quickly rotating supermassive black holes (at $1\sigma$). The
    limits disappear for lower $f_a$ because the axion cloud can
    collapse due to self-interactions before extracting a significant
    fraction of the BH's spin. The lighter regions are where the BH
    may be on a Regge trajectory and are therefore not excluded. We
    also translate the $f_a$ dependence to the quartic coupling
    $\lambda$ (right axes).}
  \label{fig:spin}
\end{figure}

Dozens of stellar and supermassive black hole spins have been measured
to date with the techniques described above. Since a black hole that
satisfies the superradiance condition loses its spin quickly on
astrophysical timescales, these measurements place limits on
previously unexplored light boson parameter space.

In fig.~\ref{fig:regge_ex} we show example regions in the BH spin-mass
plane affected by the superradiance of a QCD axion with
$\mu_a=10^{-11}\ev$.  The shaded areas correspond to the
$\ell=1,\ldots 5$ levels that satisfy the SR condition, separated by
Regge trajectories. A black hole excludes the axion mass if the SR
condition is satisfied for at least one $\ell$-level and, within
experimental error, the corresponding superradiance rate is fast
enough to grow a maximally filled cloud,
$\Gamma_{\rm{sr}}\,\tau_{\rm{bh}} \geq \log N_{\rm{max}}$.  The
relevant timescale $\tau_{\rm{bh}}$ is the shortest on which SR can be
disturbed: for stellar BHs, we use the shorter of the age and the
Eddington accretion time; for SMBHs, we use the compact-object infall
time ($\tau_{\rm{bh}}$ varies between systems so the regions shown in
fig.~\ref{fig:regge_ex} are approximate).
\begin{table*}[t!]
  \begin{center}
    \begin{tabular}{c|  c | c | c | c  |c |c|c}
      \hline
      \# &Object & Mass ($M_{\odot}$) & Spin & Age (yrs) & Period
      (days)
      &$M_{\rm{comp. \, star}}$ ($M_{\odot}$)&
      $\,\dot{M}/\dot{M}_E\,$\\ 
      \hline
      1 &M33 X-7 & $15.65\pm 1.45$ & $0.84^{+.10}_{-.10}$\cite{privatecomm2} & $3 \times 10^{6}$
      \cite{Gou:2011nq}&3.4530 \cite{pietsch2006m33} & $\gtrsim  20 $ \cite{pietsch2006m33}&$\gtrsim
      0.1$\cite{pietsch2006m33} \\
      2 &LMC X-1 & $10.91 \pm 1.4$ & $0.92^{+.06}_{-.18}$
      \cite{Gou:2009ks} & $5\times
      10^{6}$ \cite{Gou:2011nq}& 3.9092 \cite{ruhlen2011nature}&
      31.79$\pm$3.48 \cite{ruhlen2011nature} & 0.16 \cite{ruhlen2011nature}\\
      3 &\,GRO J1655-40 \,& $6.3\pm 0.5$ & $0.72^{+.16}_{-.24}$ \cite{privatecomm2} & $3.4\times 10^{8}$
      \cite{Willems:2004kk}&2.622 \cite{Willems:2004kk}& 2.3 - 4  \cite{Willems:2004kk}& $\lesssim
      0.25$ \cite{mendez1998canonical} \\
      4 &Cyg X-1 & $14.8 \pm 1.0$ & $>0.99$ \cite{2014ApJ...790...29G}& $4.8\times
      10^{6}$\cite{Wong:2011eg} &5.599829 \cite{Gou:2011nq}& 17.8  \cite{Gou:2011nq}& 0.02
      \cite{Gou:2011nq}\\
      5 &\, \,GRS1915+105 \,& $10.1\pm 0.6$ & $>0.95$ \cite{2014ApJ...796....2R,privatecomm2} & $4\times 10^{9}$
      \cite{dhawan2007kinematics}& 33.85 \cite{Steeghs:2013ksa} &0.47
      $\pm$ 0.27  \cite{Steeghs:2013ksa}& $\gtrsim
      1$ \cite{Steeghs:2013ksa}. \\
      \hline
    \end{tabular}
  \end{center}
  \caption{Stellar black holes that  set limits on light bosons
    (data
    compiled in \cite{McClintock:2013vwa} unless otherwise
    specified). Errors for masses are quoted at
    $1\sigma$,  spin limits at $2\sigma$\footnote{We thank J. Steiner
      and J. McClintock for providing the latest $2\sigma$ errors on
      the spin measurements.}.
    GRO J1655-40  has a $2\sigma$-discrepancy between continuum fitting and X-ray reflection \cite{Reynolds:2013qqa}; we
    use the continuum spin values, which are lower. GRS1915+105 has periods of unusually high
    luminosity: the spin measurement \cite{McClintock:2006xd} uses only
    data from the low-luminosity periods, when $\dot{M}/\dot{M}_E < 0.3$; in addition, we conservatively use $\tau_{bh}=\tau_{\mathrm{eddington}}/10$ to
    set the limit.} 
  \label{stellar}
\end{table*}

\begin{table}[h]
  \begin{center}
    \begin{tabular}{ c|c | c | c }
      \hline
      \#&Object & Mass ($10^6 \, M_{\odot}$) & Spin \\ 
      \hline
      1&NGC 3783 & $29.8\pm 5.4$ & $>0.88$ \\
      2&Mrk 110 & $25.1\pm 6.1$ & $>0.89$ \\ 
      3&\,MCG-6-30-15 \,& $2.9^{+.18}_{-.16}$ & $>0.98$ \\ 
      4&NGC 4051 & $1.91\pm 0.78$ & $>0.99$ \\
      \hline
    \end{tabular}
  \end{center}
  \caption{
    Supermassive black holes with reliable mass and spin measurements
    (compiled in \cite{Reynolds:2013rva,Reynolds:2013qqa}) used to set
    limits on light bosons in Fig.~\ref{fig:SMspinlim}. The mass errors
    are quoted at $1\sigma$ and the spin measurements at $90\%$ confidence. While
    many more spin measurements are available, our analysis excludes BHs
    which do not have an error estimate on the mass.}
  \label{supermassive}
\end{table}

Fig.~\ref{fig:regge_ex} also includes BH spin and mass data with
$1\sigma$ error bars, except for the lower spin limit for the two
highest spin BHs, which are quoted at $2\sigma$. The QCD axion mass
and coupling as pictured are clearly excluded by the two
fastest-spinning BHs.  Increasing the mass of the axion shifts the
affected regions to the left, relaxing the bound. Increasing the axion
self-coupling relaxes the limits as well: instead of growing to
maximum size all at once, each time the cloud reaches the critical
size $N_{\rm{bosenova}}$ it collapses, so the SR rate has to be larger
to extract the spin in the same period,
$\Gamma_{\rm{sr}}\,\tau_{\rm{bh}}\, \left({N_{\rm{bosenova}}
  }/{N_{\rm{max}}}\right) \geq \log N_{\rm{bosenova}} $. In addition,
the BH can be trapped on the Regge trajectories; if the spin and mass
of a black hole indicate that it may be on a Regge trajectory, we only
use it to place bounds if it stays there for a short time,
$\tau_{\rm{regge}} \ll \tau_{\rm{bh}}$.

We show the resulting bounds in fig.~\ref{fig:spin}. Each black hole
places a limit on a range of axion masses and each $\ell$-level leads
to the distinct lobes of the exclusion region; higher levels have
longer superradiance times and give increasingly weaker
constraints. For large axion masses, there is no measured BH light
enough to satisfy the SR condition; for axion masses too small, the
``atomic coupling'' $\alpha$ is too weak, resulting in a too-slow spin
extraction rate or too-quick mode-mixing
(section~\ref{sec:environment}) in the presence of
perturbation from the BH companion star. For strong self-interactions
($N_{\rm{bosenova}}\ll N_{\rm{max}}$), the bounds no longer apply;
this is in contrast with typical laboratory experimental limits for
axions which are relaxed
if interactions are sufficiently weak.

The bounds rely on our computation on the SR rate, and so have some
theoretical uncertainty. On the right edge of the bound, a higher SR
rate for high-$\ell$ levels would increase the exclusion while a
faster drop of the rate near the SR boundary would decrease it. The
top of the exclusion varies as the square root of bosenova size, and
logarithmically on the SR rate. At small $\alpha$ there is the
possibility of mode-mixing due to the companion star
(section~\ref{sec:environment}); the bound at light axion masses has a
$\mathcal{O}(10\%)$ uncertainty assuming $\mathcal{O}(1)$
uncertainties in SR rates and deviations of the cloud profile in the
Kerr metric from hydrogen wavefunctions.

We present more details about the stellar black holes we use to set
limits in table~\ref{stellar}. These have spins determined by both
methods, as well as precise mass measurements and an estimated age for
the BH system. Stellar BH limits are quite robust: the binary systems
are well studied, as seen from the measurements of BH properties.
These exclude the mass range $ 2\times 10^{-11} > \mu_a > 6\times
10^{-13}\ev$, corresponding to $ 3\times 10^{17} < f_a < 1\times
10^{19}\gev$ for the QCD axion: parameter space that has not been
reached with any other approach so far (but see \cite{casper} for an
experimental proposal to search for high-$f_a$ axions).

Table~\ref{supermassive} lists the masses and spins of SMBHs we use to
set limits. Their ages are unknown, but it is understood that they
accrete to reach their current mass, so the age is by definition
longer than the accretion time \cite{shankar2009self}. The dynamical
timescale $\tau_{\rm{bh}}$ is instead set by violent events. Recent
measurements indicate that a star falls into a given SMBH roughly
every $3\times 10^4$ years \cite{Khabibullin:2014iaa}. The star is
incorporated into the accretion disk; however, an in-falling BH or
neutron star could cause a large perturbation to the cloud. To
estimate the rate of such violent events, we conservatively take
$10^{-2}$ of the star infall rate, since about one in 100 stars
is a BH \cite{brown1994scenario}.

The properties of the SMBHs are less well-known, and the spin
measurements so far employ only the X-ray method
\cite{Reynolds:2013rva}, so we consider our limits
preliminary. 

As more black holes are measured with higher precision, the limits may
extend further. If, on the other hand, a light axion is nearby, the
data will begin to trace out Regge trajectories where the BH is likely
to remain for long times: in fig.~\ref{fig:regge_ex}, we expect to
find BHs only outside the SR regions or on their boundaries. This
requires a lot of progress, but can be another avenue toward axion
detection. Black holes that may be on Regge trajectories are also
candidate point sources for directed GW searches, as they may be
emitting GWs from annihilations.

 \section{Effect of the Black Hole Environment}
 \label{sec:environment}
 So far, we have considered an isolated black-hole-axion system; in
 this section, we show that this is a good approximation. The companion
 star orbits far from the black hole and the accretion disk contains a
 small fraction of the black hole mass spread over a large region; we
 will see that the perturbation due to the environment is irrelevant
 for GW signal parameter space, although it can slightly affect the bound
 derived from high-spin BHs for very large clouds, $\alpha/\ell \ll 1$.

 A non-uniform, asymmetric perturbation near the black hole can mix the
 superradiating and dumping (in-falling) levels of the cloud and cause
 part of the axion cloud to collapse. We consider a static perturbing potential
 $\delta V(\vec{\boldsymbol{r}})$: orbital frequencies of the
 companion star and accretion disk are much smaller than the axion
 energy, so the perturbation is adiabatic. The condition that at the
 horizon the energy flux is negative, i.e.  more axions are 
 extracted from the ergo-region than fall back in, is
 \begin{equation}
   \label{eq:11}
   \left|\frac{\Gamma_{\textrm{dump}}^{n'\ell
         'm'}}{\Gamma_{\textrm{sr}}^{n\ell m}} \right|^{1/2}\left|\frac{\langle
       \psi_{\textrm{dump}}^{n'\ell 'm'}|\delta V
       (\vec{\boldsymbol{r}})|\psi_{\textrm{sr}}^{{n\ell m}}
       \rangle}{\Delta E}\right| < 1,
 \end{equation}
 where $\psi_{\mathrm{sr,dump}}$ are the wave functions of the
 superradiating and dumping levels, $\Delta E$ is the energy difference
 between them, and $\Gamma_{\textrm{dump}}$ and $\Gamma_{\textrm{sr}}$
 are the analytical dumping and superradiant rates \cite{detweiler} (an
 excellent approximation at small $\alpha/\ell$). The ratio of the
 rates comes from relating the wave functions at the cloud to those at
 the horizon \cite{Arvanitaki:2010sy} and scales as a high power of
 $\alpha$,
 \begin{equation}
   \label{eq:13}
   {\Gamma_{\textrm{dump}}^{n'\ell'm'}}/{\Gamma_{\textrm{sr}}^{n\ell
       m}}\propto\alpha^{4(\ell'-\ell)}\quad (\alpha/\ell \ll 1),
 \end{equation}
 with a weaker dependence on $m$.

 The energy differences $\Delta E$ are $\ma (\alpha/n)^2$
 between levels with $\Delta n \neq 0$, with $\sim \ma
 (\alpha/\ell)^4$ ``fine'' level-splitting from relativistic corrections
 (if $\Delta \ell \neq 0$) and $\sim \ma(\alpha/\ell)^5$ ``hyperfine''
 splitting from spin-orbit coupling (if $\Delta m \neq 0$),
 \begin{equation}
   \Delta H_{\rm{s.o.}} = \boldsymbol{\mu}_{\rm{axion-orbit}} \cdot
   \boldsymbol{B}_{\rm{BH}} \sim \frac{a_*}{ r_g}
   \left(\frac{\alpha}{\ell}\right)^6 ,
 \end{equation}
 and higher order corrections to the Newtonian potential from the black
 hole. 

 \subsection{Companion Star}

 The observed stellar BHs are in binaries with companion stars, the
 masses and especially orbital periods of which are known to great
 precision (Table~\ref{stellar}).  For the systems used in setting
 limits, the companions have mass $M_*$ of order the BH
 mass and orbit at a distance $R\sim 10^6 \, r_g$ from the BH. To
 compute the effect of the companion, we decompose its gravitational
 potential into multipoles; schematically,
 \begin{equation}
   \label{eq:10}
   \frac{\delta V}{\mu_a} \sim \frac{M_{*}}{M}\frac{r_g}{R}\left(1 +
     \frac{r_c}{R} \,Y_{1,m} +\frac{r_c^2}{R^2} \,Y_{2,m}+\dots\right).
 \end{equation}
 The leading contribution comes from the dipole ($\ell' = \ell-1$) term
 of order $r_c^2/R^2\ll 1$; the resulting mixing does not affect
 superradiant growth if
 \begin{equation}
   \l(\frac{\alpha}{\ell}\r) \gtrsim (0.05)
   \l( \frac{M_{*}}{10 M_{\odot}}\r)^{\frac{1}{8}}\l(\frac{M}{10 M_{\odot}}\r)^{\frac{1}{24}}
   \l(\frac{\mathrm{day}}{T}\r)^{\frac{1}{6}},
   \label{eq:alphalim}
 \end{equation}
 where $T$ is the orbital period of the companion star and $M$ is the black hole mass.  Higher
 multipoles mix with other $\ell'\neq \ell$ modes and give a comparable
 bound, while mixing with $\Delta \ell = 0, m' \neq m$ modes gives a
 weaker bound since the ratio of dumping to superradiance rates for
 these modes is smaller.  Thus, a typical companion star may disrupt
 superradiance only if the axion-black hole coupling is small,
 $\alpha/\ell < 0.05$; the bound has very weak dependence on the black
 hole binary parameters. We take this constraint into account in
 setting limits on axion parameter space. For our GW signal estimates,
 the bound on $\alpha/\ell$ is irrelevant: the weak coupling 
 produces signals too small to observe, and signals are just as likely
 to come from the over $50\%$ of BHs which are not in binaries
 \cite{shakura1973black}.

 \subsection{Accretion Disk}

 Stellar BHs in binaries and supermassive BHs are surrounded
 by accretion disks which extend out from the innermost stable orbit.
 For stellar black holes, the accretion disk extends to
 the companion star at $10^{6}~r_g$, while for a $M = 10^8\msun$ black
 hole, the disk ends at $\sim 10^3~r_g$ \cite{acdiskbook}. Although the disk can
 spatially overlap with the axion cloud, it does not source a large
 perturbing potential since the mass in the disk is very small compared
 to the BH mass. To compute the effect on the axion cloud, we use the
 middle region of the thin-disk model for the surface mass-density (for $r \gg r_g$)
 \cite{shakura1973black}:
 \begin{equation}
   \!\!\sigma(r)\!\simeq\!
   \frac{10^{-17} M}{r_g^{2}}\!\left(\!\frac{0.01}{\alpha_{\rm{disk}}}\!\right)^{\frac{4}{5}}\!\left(\!\frac{L}{L_{\rm{edd}}}\!\right)^{\frac{3}{5}}\!\left(\!\frac{M}{\msun}\!\right)^{\frac{6}{5}}
   \!\left(\!\frac{r}{r_g}\!\right)^{-\frac{3}{5}}\!\!,
   \label{eq:disk}
 \end{equation}
 where $L \lesssim L_{\rm{edd}}$ is the disk luminosity
 and $\alpha_{\rm{disk}}= 0.01$--$0.1$ characterizes the disk properties;
 we conservatively use $\alpha_{\rm{disk}} = 0.01$ and $L =
 L_{\rm{edd}}$ in our estimates.  Even maximally accreting
 disks contain a tiny amount of mass, and the effect on the axion cloud is
 further suppressed by the small height of the disk, $z(r) \simeq 10^{-2}
 r$ \cite{shakura1973black}.

 To find a bound on $\alpha/\ell$, we numerically evaluate
 eq.~\eqref{eq:11} using potential perturbation sourced by
 eq.~\eqref{eq:disk}. Of course only non-uniformities in the disk
 contribute to the mixing of different levels and the fractional mass
 in a given mode is even smaller, but to avoid model-dependence of the
 disk substructure, we conservatively take the entire mass of the disk
 to be concentrated in a spherical harmonics mode that induces mixing
 of a given SR mode with the fastest-dumping mode. The scenario with
 the biggest ratios of $\Gamma_{\rm{dump}}/\Gamma_{\rm{sr}}$ (thus
 giving the most stringent bounds) is mixing with the rapidly decaying
 $\ell' = m' = 0$ mode, followed by the $m' =-m, \ell' =\ell$ mode.

 \begin{figure}[t]
   \centering
   \includegraphics[width=.9\linewidth]{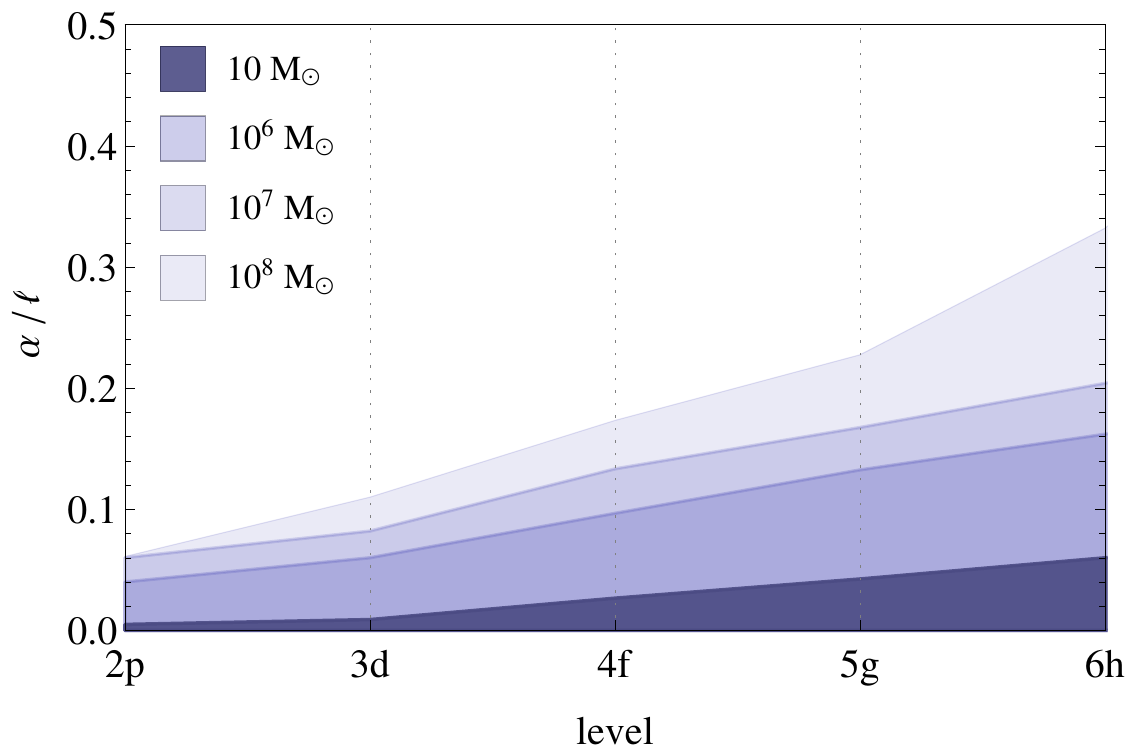}
   \caption{Values of $\alpha/\ell$ for different levels ($x$-axis) affected by the gravitational
     potential of the accretion disk. For each level, we assume the entire mass
     distribution conspires to form the angular mode causing the most
     dangerous mixing between superradiant and dumping modes. For
     comparison, the affected range from companion stars for stellar BHs
     is $\alpha/\ell<0.05$. }
   \label{fig:accretion_bound}
 \end{figure}

 In fig.~\ref{fig:accretion_bound} we show the resulting bound for the
 axion clouds of different levels. We emphasize that these are upper
 bounds on the region affected by the accretion disk. We see that for
 stellar BHs, the disk constraint on $\alpha/\ell$ is weaker than that
 from the companion star. The disks of SMBHs are fractionally smaller
 in extent, but disk density grows quickly with the BH mass, so the
 effect of accretion disks around SMBHs is relatively larger. The
 higher-$\ell$ levels are more affected by the accretion disk since
 their superradiance rates are increasingly smaller than the dumping
 rate of the $\ell ' = m' = 0$ level. However, the superradiance time
 for these levels becomes long and limited by other factors, so they
 are not relevant for the limits or signals we consider; in agreement
 with numerical results in \cite{Brito:2014wla} the relevant time
 scale is $\sim 100$ SR times. In sum, even with very conservative
 assumptions, the effect of the accretion disk on the axion cloud does
 not constrain parameter space where the effect of superradiance is
 relevant.

\section{Summary}
\label{sec:conclusion}

\begin{table}[h]
  \begin{center}
    \begin{tabular}{  l |c | c  c   c |c }
      \hline
      GW signal source&$M$& $~{N}_{\mathrm{low}}~$ & $~{N}_{\mathrm{re}}~$&
      $~{N}_{\mathrm{high}}~$&$~f \,(\mathrm{Hz})$\\ 
      \hline\hline
      $6g\!\rightarrow\! 5g$ transition& $10\, M_{\odot}$ &
      $0.01$ & $0.1$
      & $0.5$ & $30$\\
      $2p$ annihilation & $30\, M_{\odot}^*$ & $30$&
      $300$&$2000$&$600$\\
      \hline
      $6g\!\rightarrow\! 5g$ transition& $10^6\, M_{\odot}$&$10^{-4}$&  $0.1$&$1$& $3\times 10^{-4}$\\
      $2p$ annihilation&  $10^6\, M_{\odot}$&$10^{-3}$&$10$ &$60$&$~2\times10^{-2}$\\
      $2p$ annihilation& $10^7\, M_{\odot}$ &$10^{-3}$&$20$ &$300$&$~2\times10^{-3}$\\
      \hline
    \end{tabular}
  \end{center}
  \caption{Estimated event rates for transition and annihilation signals
    at aLIGO and eLISA detectors , if there exists an axion with
    mass that falls in the sensitive bands of the detectors. These values are simplified following the
    LIGO event estimation method \cite{Abadie:2010cf} by assuming
    that all stellar (supermassive) black holes have mass $10\msun$
    ($10^6/10^7\msun$), except for the $30\msun$ value: we take $10^{-3}$ of
    stellar black holes to have this mass. We take into account spin and distance
    distributions of black holes.}
  \label{tab:detectionrates}
\end{table}

Advanced LIGO and VIRGO will soon start their science runs, and with the help of
the process known as black hole superradiance, they will be the first
experiments with potential to discover the QCD axion with decay
constants above the GUT scale. Superradiance fills levels of the
gravitational ``atom'' formed by the black hole and the axion, creating
a macroscopic ``cloud'' of axions with large occupation
number. Advanced LIGO will be sensitive to gravitational wave
radiation from axion annihilations (for Planck-scale QCD axions) and
axion transitions between levels (for GUT-scale QCD axions). Advanced
VIRGO has similar sensitivity. 

Both signals are monochromatic and last dozens of years or more, with
time-dependent intensity and frequency drift of $10^{-11}$ Hz/s or
less. These signals are distinct from monochromatic radiation from
rapidly spinning asymmetric neutron stars, which have negative
frequency drifts following the stars' spin-down rate, while both
transition and annihilation signals have frequency drifts
anti-correlated in time with the signal intensity.

We extrapolate the current LIGO run's monochromatic search
\cite{aasi2013einstein} to aLIGO sensitivities, and we estimate that
aLIGO should expect up to $\mathcal O(1)$ persistent events from axion
transitions around stellar black holes. Since the signal lasts longer
than the duration of the experiment, this event estimate for
transitions could be interpreted as the probability of observing the
axion at aLIGO.

For axion annihilations, the optimal reach of aLIGO corresponds to BHs
near $30 M_{\odot}$ in mass for which very little is known. If we
extrapolate measured stellar BH distributions to higher masses, we
expect thousands of events at aLIGO for axions around $6\times
10^{-13}\ev$ in mass.  Part of the parameter space where annihilations
give appreciable event rates is disfavored by BH spin measurements,
and in the event that aLIGO does not observe a signal, the resulting
limits would provide a cross-check on the spin measurement
constraints.

Both axion transitions and annihilations provide the opportunity to
discover the QCD axion within a few years' time through its
gravitational coupling. Future GW observatories such as the Einstein
Telescope may reach a factor of $10$ further in sensitivity than
aLIGO \cite{punturo2010einstein} and detect a factor
of $10$ or more events than our Advanced LIGO estimates.

A focused search on point sources may also be promising. In
particular, annihilation signals from nearby fast-spinning BHs such as
those in table~\ref{stellar} can probe axion masses of $\ma >
10^{-11}\ev$; however, given the age of the BHs and the relatively
fast timescales of superradiance and annihilation, we need to be lucky
to see such signals. Other point sources worth studying are BHs newly
formed after supernovae and binary mergers; the formation events of
these BHs can be correlated with the annihilation signals that develop
after superradiance has had the time to evolve.

As we can see from table~\ref{tab:detectionrates} and
fig.~\ref{fig:lisaevents}, the prospect for discovery of much lighter
axions (between $10^{-19}$ and $10^{-16}$ eV) is also promising
through the annihilation signal around supermassive BHs at future
lower-frequency GW observatories such as eLISA and AGIS. We expect to
see up to $10^3$ annihilation events, with a large uncertainty. Level
transition signals from supermassive black holes would mostly come
from BHs with masses below $10^5 M_{\odot}$. Very little is known
about BHs in this mass range, but our estimate in
table~\ref{tab:detectionrates} shows that events can also be
observed. Our SMBH event rates have a very large uncertainty due to
unknown spin distributions and astrophysical uncertainties in the SMBH
environment.

One signature we have not explored in detail is GW emission from
``bosenovae'', the collapse of the cloud under its self-interaction,
relevant for bosons with self-interaction stronger than that
of the QCD axion. Rates for bosenova events are difficult to estimate
because the shape and frequency of the signal are sensitive to the
dynamics of cloud collapse. This is a particularly interesting avenue
for future numerical studies, since the signal has promising amplitude and
a distinctive time profile.

The range of axion masses probed by GW detectors is already
constrained by black hole spin measurements. For supermassive BHs, these
measurements are less reliable given the uncertainties in the spin
measurement method as well as the infall rates of compact objects. For
stellar BHs, spin measurements are confirmed by two independent
techniques, and the environment of the BH is well-known in the
relevant cases. This makes the exclusion of $6\times 10^{-13} \ev <
\ma < 2 \times 10^{-11} \ev$ 
quite robust. This bound is taken into account in our conclusions for
the discovery potential of aLIGO. 

Astrophysical black hole superradiance diagnoses the presence of light axions
in the theory independently of their cosmic evolution and abundance.
We focus on the QCD axion in this paper, but the above discussion can be generalized
to all spin-0 bosons with weak self-coupling since gravity is the only
interaction required. Effects we discussed can also be extended to
light spin-1 particles, although further study, in particular of their
superradiance rate, is needed.

Throughout this paper, we make several assumptions regarding BH mass
and spin distributions; these assumptions lead to the range of event
rate expectations for the different signatures presented in
figs.~\ref{fig:aLIGOreach}, \ref{fig:aLIGOreach_ann},
\ref{fig:lisaevents}, and table~\ref{tab:detectionrates}. Most of the
extended range come from the uncertainties in the expected spin
distributions; this is especially true for SMBHs since their spin
depends on their integrated history which is unknown. Future
measurements will shed light on these distributions and will narrow
the range presented. Uncertainties due to BH mass distributions are
less significant; using different BH mass distributions presented in
the literature makes $\mathcal{O}(1)$ difference in the event rate, an
uncertainty dominated by large uncertainties in the spin
distribution. For stellar BHs annihilation signals, the range of
accessible axion masses depends on the tail of the BH mass
distribution; we model this uncertainty by imposing a range of upper
bounds on the BH mass.

We also use several approximations in our GW power
calculations; we expect that our analytic approximations for the GW power
calculation are sufficient at this stage. As shown in numerical
studies \cite{Yoshino:2013ofa}, relativistic effects are important for
annihilation signals, but our use of the flat space approximation
 consistently underestimates the power of the signal compared
to numerical results even at large values of $\alpha / \ell$. We thus
consider this approach conservative and appropriate given the large
astrophysical uncertainties. For transitions, the signal arises from
quadrupole radiation, and deviations from our estimates at the large
$\alpha / \ell$ regime should be at the $\mathcal{O}(1)$ level. Full
numerical studies to take into account higher-order corrections and
non-linear effects would be required for further analysis and in the
event that a candidate signal is discovered. Numerical studies are
also indispensable in the study of the bosenova effect.

The work presented above shows that the imminent discovery of
gravitational waves not only gives us a chance to study the properties
of black holes but also has the potential to diagnose the presence of
new particles. The prospects for discovery are exciting, and our work
may only be scratching the surface of the rich phenomena of black hole
superradiance.



\SkipTocEntry\acknowledgments{}
\noindent We are grateful to Tom Abel, Kipp Cannon, Savas Dimopoulos,
Sergei Dubovsky, Peter Graham, Junwu Huang, John March-Russell, Joshua
Smith, Ken Van Tilburg, and Natalia Toro for many useful comments and
discussions. We especially thank Robert Wagoner for his patience and
enthusiasm in helping us learn about black holes. AA would like to
express her appreciation to Gabriela Gonzalez and Xavier Siemens for
extremely useful discussions. AA also extends her gratitude to Luis
Lehner both for physics discussions and for introducing her to LIGO
experimentalists. We thank the CERN Theory Group for their hospitality
during the completion of this work. This research was supported in
part by Perimeter Institute for Theoretical Physics. Research at
Perimeter Institute is supported by the Government of Canada through
Industry Canada and by the Province of Ontario through the Ministry of
Economic Development \& Innovation. This work was supported in part by
ERC grant BSMOXFORD no. 228169.  MB is supported in part by the
Stanford Graduate Fellowship. XH is supported in part by the NSF
Graduate Research Fellowship under Grant No. DGE-1147470.

\appendix

\section{Gravitational Wave Power Calculation}
\label{sec:app_GW_power}

To calculate the transition rate $\Gamma_t$ and annihilation rate
$\Gamma_{a}$ (sections \ref{sec:transition} and
\ref{sec:annihilation}) we make two approximations here as in
\cite{Arvanitaki:2010sy}: we use the flat space formula for the
gravitational wave flux and the hydrogen wave functions for the axion
wavefunctions, both of which are valid since the cloud is
localized far away from the BH horizon. Further numerical study would be of interest.

For source stress-energy tensor decomposed as
\begin{equation}
  \label{eq:6}
  T_{\mu\nu} = \sum_{\omega}e^{-i\omega t}T_{\mu\nu}(\omega,\vecc{x}) + \cc,
\end{equation}
the power per solid angle emitted in a direction $\vechat{k}$ is
\cite{weinbergGR},
\begin{equation}
  \label{eq:4}
  \frac{\mathrm{d}P}{\mathrm{d}\Omega}=\sum_{\omega}\frac{G_N\omega^2}{\pi}\Lambda_{ijlm}T^{ij}(\omega,\vecc{k})\,T^{lm*}(\omega,\vecc{k})
\end{equation}
where $\vecc{k}^2 = \omega^2$,
\begin{equation}
  \Lambda_{ijlm} = P_{mi}P_{jl}-\frac{1}{2}P_{ji}P_{ml},\quad P_{ij} = \delta_{ij}-\frac{k_{i}k_j}{k^2},
\end{equation}
and $T_{ij}$ is the cartesian spatial component of the Fourier
transform
\begin{equation}
  T_{\mu\nu}(\omega,\vecc{k}) = \int d^3\vecc{x}\, e^{-i\vecc{k}\cdot\vecc{x}}\,T_{\mu\nu}(\omega,\vecc{x}).
\end{equation}
We use indices $i,j,\dots$ to denote cartesian spatial coordinates for
the rest of this section.

We start with the classical axion field,
\begin{equation}
  \phi(r,\theta,\varphi) = \sum_{n,l,m}e^{-i \omega_{n} t}\sqrt{\frac{N_{nlm}}{2 \ma}}\Psi_{nlm}(r,\theta,\varphi)+\rm{c.c.}
\end{equation}
where $N_{nlm}$ is the occupation number and $\Psi_{nlm}$ is the
normalized hydrogen wave function for the $(n,l,m)$ level with energy
$\omega_{n}$ (eq.~\eqref{eq:energies}), with stress energy tensor
\begin{equation}
  T_{\mu\nu} = \nabla_\mu \phi \nabla_\nu \phi
  -g_{\mu\nu}\left(\frac{1}{2}\nabla_\rho\phi\nabla^\rho\phi+V(\phi)\right).
\end{equation}
In the Minkowski metric the spatial part of the second term is
proportional to $\delta_{ij}$ and does not contribute to
eq.~\eqref{eq:4}; therefore the cartesian spatial component is
\begin{equation}
  \label{eq:5}
  T_{ij}(r,\theta,\varphi) = \nabla_{i}\phi\nabla_j\phi,
\end{equation}
where
\begin{equation}
  \nabla_{i} \!=\! \frac{1}{r}\!\left(\!
    \begin{array}{ccc}
      r \cos  \varphi  \sin  \theta  & \cos  \theta  \cos  \varphi  & -\csc  \theta  \sin  \varphi  \\
      r \sin  \theta  \sin  \varphi  & \cos  \theta  \sin  \varphi  & \cos  \varphi  \csc  \theta  \\
      r \cos  \theta  & -\sin  \theta  & 0 \\
    \end{array} \!\!\right)\! \left(\! \begin{array}{c}
      \partial_{r}  \\
      \partial_{\theta}  \\
      \partial_{\varphi}  \end{array} \!\right)
\end{equation}

\begin{table}
  \centering
  \begin{tabular}{c|c}
    \small{Level}& $\mathrm{d}P/\mathrm{d}\Omega \,N^{-2}$\\
    \hhline{=|=}
    $2p$ & $\frac{\alpha ^{18} G_N \left(6 \alpha ^3+40 \alpha- 3
        \left(\alpha ^2+4
        \right)^2\tan^{-1}(2/\alpha)\right)^2 \left(28 \cos 2 \theta _k +\cos4 \theta_k  +35\right)}{2^{24} \pi   \left(
        \alpha ^2+4 \right)^4 r_g^4}$\rule{0pt}{3.5ex}\\[3ex]
    $3d$ & $\frac{\alpha ^{20} G_N \sin^4\theta_k
      (28  \cos 2 \theta_k + \cos 4 \theta _k +35)}{2^4 3^{16} \pi  r_g^4
    } + \dots$\\[3ex]
    $4f$ & $\frac{\alpha ^{24} G_N \sin^8\theta_k
      (28  \cos 2 \theta_k + \cos 4 \theta _k +35)}{5^{-2}2^{24} \pi  r_g^4
    } + \dots$\\
  \end{tabular}
  \caption{Analytic expressions for GW power from axion annihilations,
    with $m = l$. For brevity, we expand in $\alpha$ for higher 
    $l$ levels, though the first few higher-order terms are comparable
    in magnitude.}
  \label{tab:ann_power}
\end{table}

\begin{table}
  \centering
  \begin{tabular}{c|c}
    Level  & $\mathrm{d}P/\mathrm{d}\Omega \,N_{nlm}^{-1}N_{n'l'm'}^{-1}$\\[0.6ex]
    \hhline{=|=}
    $6g\rightarrow 5g$\,\, & $\frac{ 2^{28} 3^4 5^5 \alpha ^{12} G_N
      \sin^4\theta _k }{ 11^{22}\pi  r_g^4}+\dots$\rule{0pt}{3.5ex} \\[3ex]
    $7h\rightarrow 6h$ & $\frac{2^{31} 3^7 5^2 7^6 \alpha ^{12} G_N \sin ^4\theta_k}{13^{26}\pi  r_g{}^4 }+\dots$\\[3ex]
    $5f\rightarrow 4f$ & $\frac{ 2^{22} 5^2 \alpha ^{12} G_N
      \sin^4\theta _k }{ 3^{34}\pi  r_g^4}+\dots$ \\
  \end{tabular}
  \caption{Analytic expressions for GW power from transitions. Higher order terms in $\alpha$ are
    smaller by a factor of $10$ or more. }
  \label{tab:trans_power}
\end{table}

Decomposing \eqref{eq:5}, we identify terms in the expansions
\begin{align}
  \label{eq:7}
  &\left.T_{ij}(r,\theta,\varphi)\right|_{\rm{ann}} \equiv
  T_{ij}(\omega_{n}+\omega'_{n},r,\theta,\varphi)\\
  &=
  e^{i\left(\omega_{n}+\omega'_{n}\right)t}\frac{(N_{nlm}N_{n'l'm'})^{\frac{1}{2}}}{2\ma}\left(\nabla_i\Psi_{nlm}\right)\left(\nabla_j\Psi_{n'l'm'}\right)\nonumber
\end{align}
as the GW source from annihilation of two axions from levels $(n,l,m)$ and 
$(n',l',m')$, and
\begin{align}
  \label{eq:8}
  & \left.T_{ij}(r,\theta,\varphi)\right|_{\rm{trans}} \equiv
  T_{ij}(\omega_{n}-\omega'_{n},r,\theta,\varphi)\\
  &=
  e^{i\left(\omega_{n}-\omega'_{n}\right)t}\frac{(N_{nlm}N_{n'l'm'})^{\frac{1}{2}}}{2\ma}\left(\nabla_i\Psi_{nlm}\right)\left(\nabla_j\Psi^*_{n'l'm'}\right)
  + \cc\nonumber
\end{align}
as the GW source from transition from levels $(n,l,m)$ to $(n',l',m')$.
To calculate the GW power, we use eqs. \eqref{eq:7} \& \eqref{eq:8} and Fourier
transform to momentum space,
\begin{align}
  \label{eq:9}
  T_{ij}(\omega,\theta_k,\varphi_{k}) &= \int dr \, d\theta\, d\varphi \,\, r^2\sin\theta\\
  \times \sum_{l,m} 4\pi  &(-i)^l j_l(\omega r) Y_{lm}^*(\theta,\varphi)Y_{lm}(\theta_k,\varphi_k)T_{ij}(\omega,r,\theta,\varphi),\nonumber
\end{align}
where $j_l$ are the spherical Bessel functions and $Y_{lm}$ are the
spherical harmonics.

We then plug eq.~\eqref{eq:9} into eq.~\eqref{eq:4} to calculate the
differential power. To calculate the annihilation and transition
rates, we use
\begin{equation}
  \!\Gamma_a \equiv \frac{\int\! \mathrm{d}\Omega \,\mathrm{d}P/\mathrm{d}\Omega|_{\rm{ann}}}{2\omega N^2}, \,\,\text{and}\,\,
  \Gamma_t \equiv \frac{\int\! \mathrm{d}\Omega \,\mathrm{d}P/\mathrm{d}\Omega|_{\rm{tr}}}{\omega N N'},
\end{equation}
respectively.  We list the
differential power for the most relevant annihilations and transitions
in tables~\ref{tab:ann_power} and \ref{tab:trans_power}.

For the $2p$ level, we find a cancellation in the leading $\alpha$
term because our calculation is performed in the flat space
approximation. The leading $\alpha$ term is restored if we take into
account the first-order corrections due to the BH gravitational
potential, which agrees with the calculation in the Schwarzschild
background in \cite{Brito:2014wla} and the numerical result in
\cite{Yoshino:2013ofa}; we thank the authors of \cite{Yoshino:2013ofa}
for clarifying this issue. Our calculation, which we adopt throughout
this paper, should thus be considered as a lower bound on the GW
production.

\section{Reach of GW experiments}
\label{sec:app_exp_reach}
For each mass $\ma$, we calculate the reach to a black hole of mass
$M$ that optimizes the signal strength and with spin $a_*=0.9$ or
$a_*=0.99$.  In each plot, the top axis shows the corresponding GW
frequency. We assume a maximally filled axion cloud for annihilations
and a peak transition strain for $6g\rightarrow 5g$ transitions.

We plot the reach of aLIGO \cite{Aasi:2013wya}, eLISA\cite{LISA}, and AGIS
\cite{Dimopoulos:2008sv} based on their design strain sensitivities, using 121 segments
of 250 hr. integration time and a trials factor of 10 as described in
section~\ref{sec:gw_signals}. We extrapolate the AGIS sensitivity to
lower frequencies (dashed curve) assuming the same
frequency-dependence of the noise floor as that at mHz.

We indicate with vertical lines when the optimal BH masses are heavier
than $100 \msun$ and lighter than $3 \msun$ (stellar BHs), or heavier
than $10^9 \msun$ and lighter than $10^6 \msun$ (SMBHs). Outside of
this range, there are few observations to guide estimates of BH
mass distributions.

The shaded regions are disfavored by the observation of rapidly
spinning black holes for bosons with coupling equal to that of the QCD
axion (light gray) or stronger (dark gray) (see section
\ref{sec:spin_limit}).

\section{Event Rate Calculation}
\label{sec:app_bh_properties}

We calculate the event rates by incorporating the various astrophysical
distributions:
\begin{align}
  \text{\# of Events} \equiv& \int_{0}^1\!
  \mathrm{d}a_{*}P(a*)\int_{0}^\infty\! \mathrm{d}r P(r)
  \int_{0}^{M_{\rm{max}}}\! \mathrm{d}M P(M)\,\times \nonumber\\
  &{\tau_{\rm{sig}}(M, a_*,r)}{\,\times\,\text{BHFR}\,}
\label{eq:eventnum}
\end{align}
where BHFR is the black hole formation rate, $P(a_*)$, $P(M)$, $P(r)$
are the normalized probability distributions to find a black hole with
spin $a_*$, mass $M$, and distance $r$ away, and
$\tau_{\rm{sig}}(M,a_*,r)$ is the duration for which the GW signal
from a BH with mass $M$, spin $a_*$, and distance $r$ away is above
the noise threshold of the detector.
\vspace{-.2cm}

\SkipTocEntry\subsection{Stellar Black Hole Distributions}
\textit{Mass distribution}: We quantify the probability of finding a
black hole of a certain mass by using an exponential fit to current
data, which is found to be the best fit out of 5 functional forms
(table 7 in \cite{Farr:2010tu}):
\begin{equation}
P(M) = M_0^{-1} e^{\frac{M_{\rm{min}}}{M_{0}}} e^{-\frac{M}{M_{0}}}.
\label{eq:massdistr}
\end{equation}
Here $M_{\rm{min}} = 5.3^{+0.9}_{-1.2} \msun$ is the minimum BH mass
which can be formed from stellar collapse and
$M_0=4.7^{+3.2}_{-1.9} \msun$ sets the width of the distribution
(table 5 in \cite{Farr:2010tu}). We verified that the uncertainty in
the range of values we use for the exponential fit is larger than the
variation with respect to the second best-fit function (``double
gaussian'').

In computing the event rates for transitions, we use
$M_{\rm{min}} = 4.1 \msun$, within the experimental fit and above
theoretical maximum of neutron star mass, $\lesssim 2.9 \msun$
\cite{maximumNeutronStarMass}; lowering $M_{\rm{min}}$ shifts the
event distribution to higher axion masses. Larger values of $M_0$ give wider distributions.
We use the following three values for $M_0$ parameter for the three
bands in fig.~\ref{fig:aLIGOreach}:
\begin{itemize}
\item Narrow: $M_0 = 2.8\msun$
\item Intermediate: $M_0 = 4.7\msun$
\item Wide: $M_0 = 7.9\msun $
\end{itemize}
We expect the narrower distribution is more accurate for young BHs
because some larger masses are attained by accretion.

In annihilation event rates (fig.~\ref{fig:aLIGOreach_ann}), we vary
$M_{0}$ as part of our pessimistic and optimistic estimates,
respectively, since the aLIGO event rate is sensitive to the
exponential tail of massive BHs. We plot three bands in
fig.~\ref{fig:aLIGOreach_ann} corresponding to different hard cutoff
$M_{\rm{max}}$ to further account for this uncertainty: 

\begin{figure}[h]
  \includegraphics[trim = 0mm 0mm 0mm 0mm, clip, width
  = 0.95\linewidth]{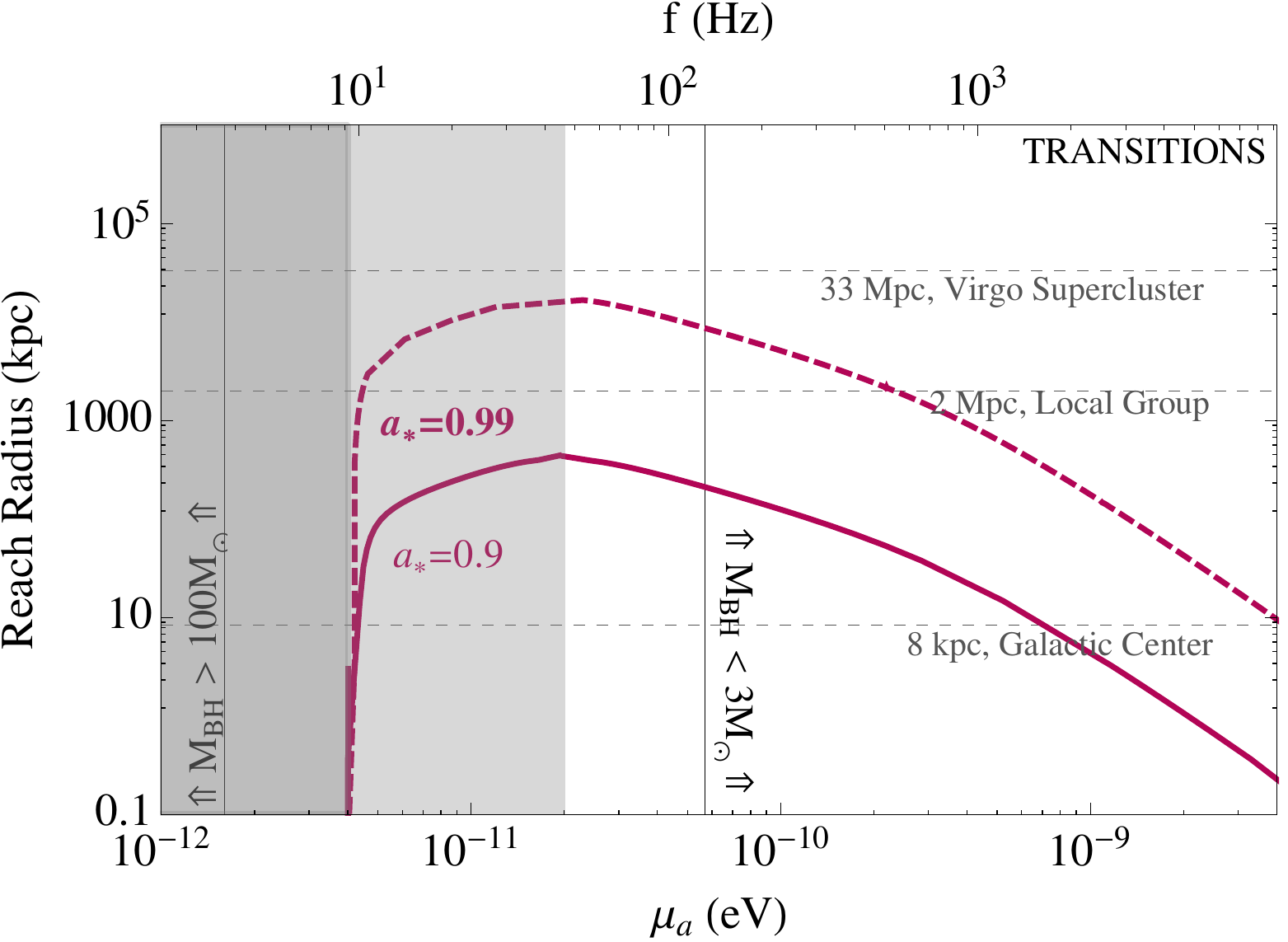}
  \vspace{-4mm}
  \caption{ aLIGO reach for a black hole-axion ``atom''
    currently undergoing the $6g\rightarrow 5g$ transition as a
    function of the axion mass at the maximum rate for a black hole
    spin of $a_*=0.9$ (solid) and $a_*=0.99$ (dashed).
  }
  \label{fig:treach} 
  \vspace{-3mm}
\end{figure}

\begin{figure}[h]
  \begin{center}
    \includegraphics[trim = 0mm 0mm 0mm 0mm, clip, width
    = 0.92\linewidth]{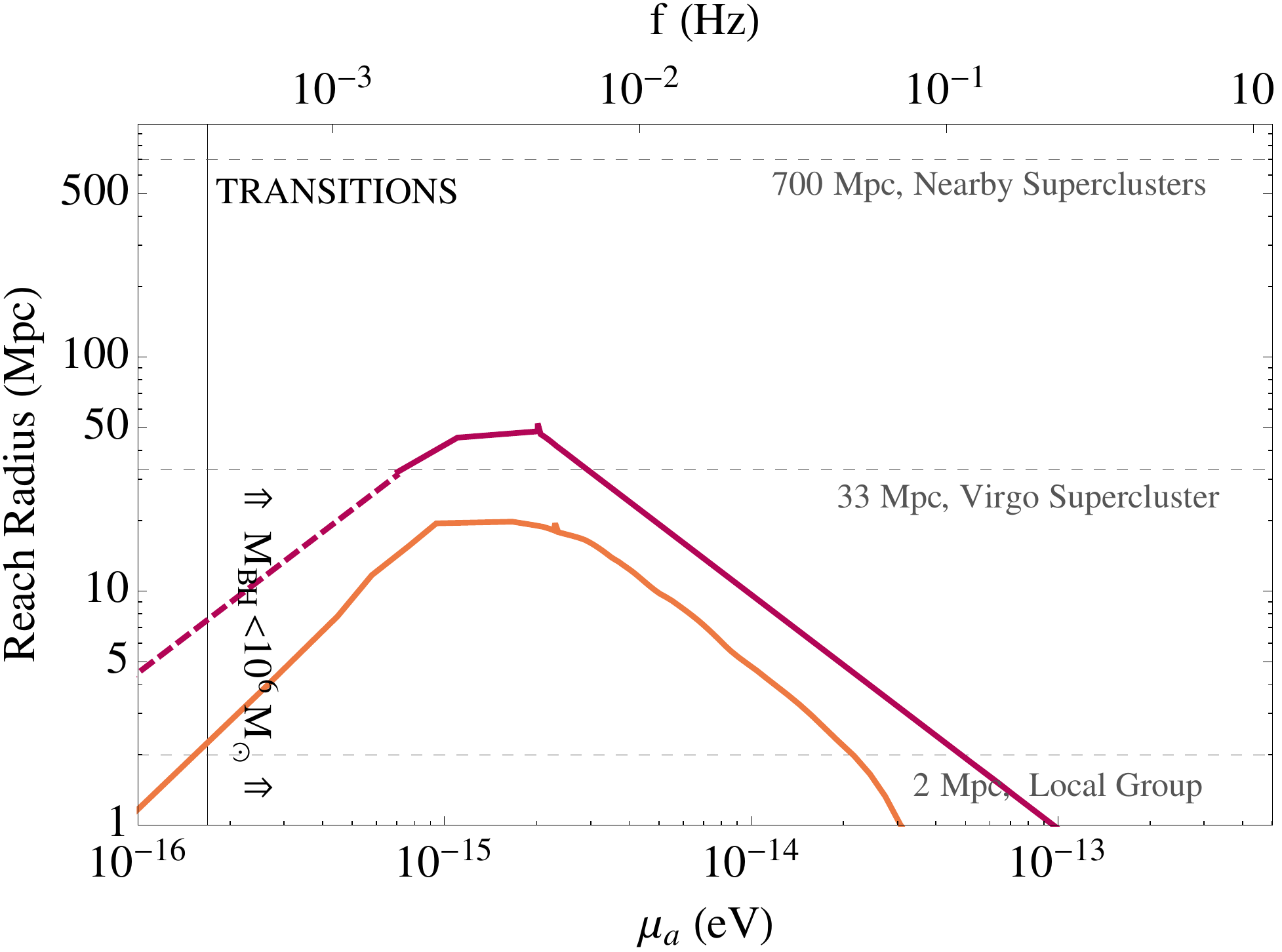}
    \vspace{-4mm}
    \caption{ AGIS (red/dark gray) and eLISA (orange/light gray) reach
      as a function of the axion mass for a black hole-axion ``atom''
      currently undergoing the $6g\rightarrow 5g$ transition at the
      maximum rate for BHs with spin $a_* = 0.99$. For $a_* = 0.9$ the
      reach is decreased by a constant factor $\sim 50$ as in
      fig.~\ref{fig:treach}, making the signal from extragalactic BHs barely
      observable. } \label{fig:treach_sm} \vspace{-5mm}
  \end{center}
\end{figure}

\begin{figure}[h]
  \begin{center}
    \includegraphics[trim = 0mm 0mm 0mm 0mm, clip, width
    =0.95\linewidth]{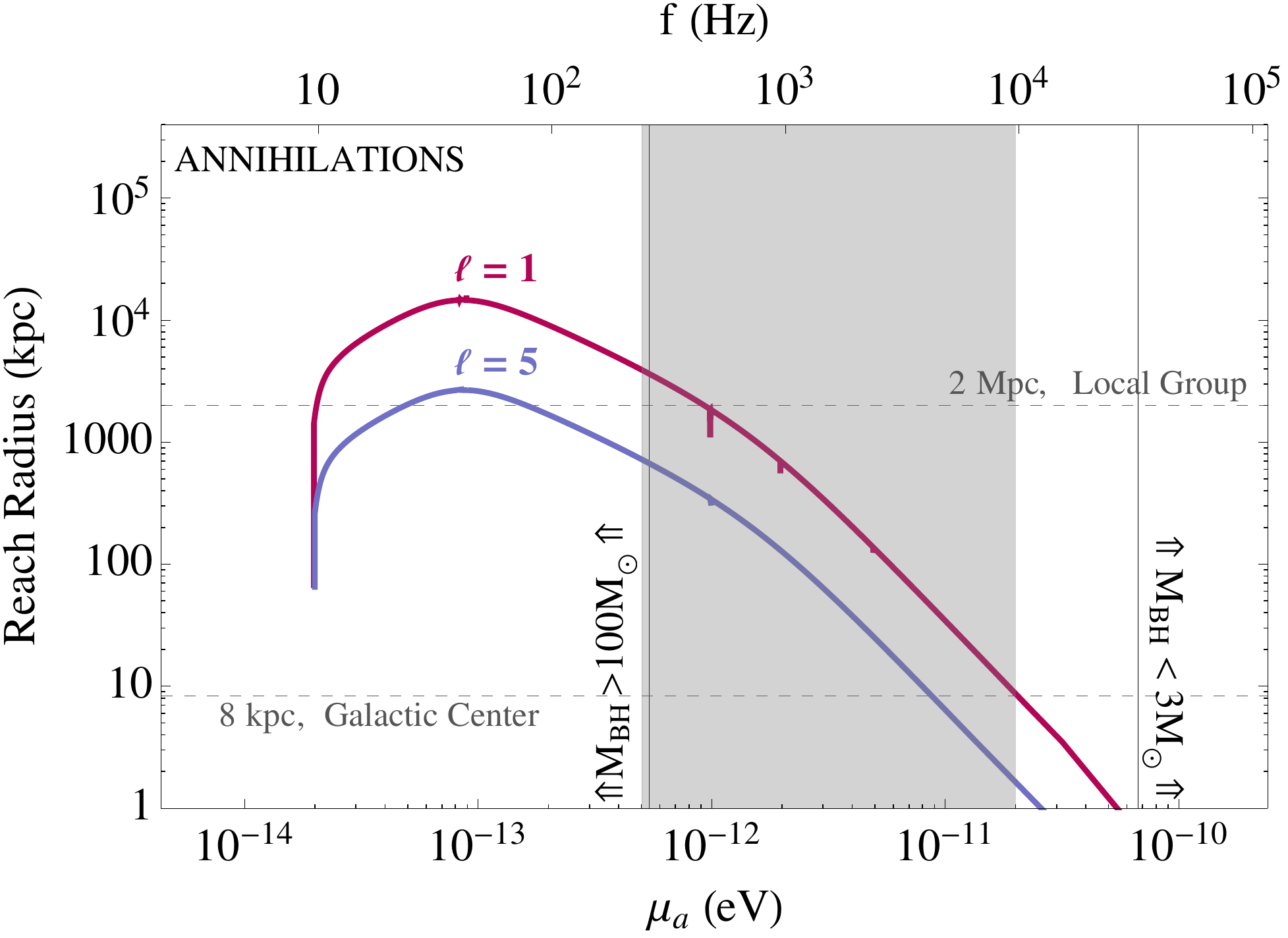}
    \vspace{-4mm}
    \caption{ aLIGO reach as a function of the axion mass for an axion
      cloud currently undergoing annihilations at the maximum rate
      $a_* = 0.9$ for  ($\ell=1$ and
      $5$, $n=\ell+1$).
    } \label{fig:areach} \vspace{-5mm}
  \end{center}
\end{figure}

\begin{figure}[h!]
  \begin{center}
    \includegraphics[trim = 0mm 0mm 0mm 0mm, clip, width
    =.95\linewidth]{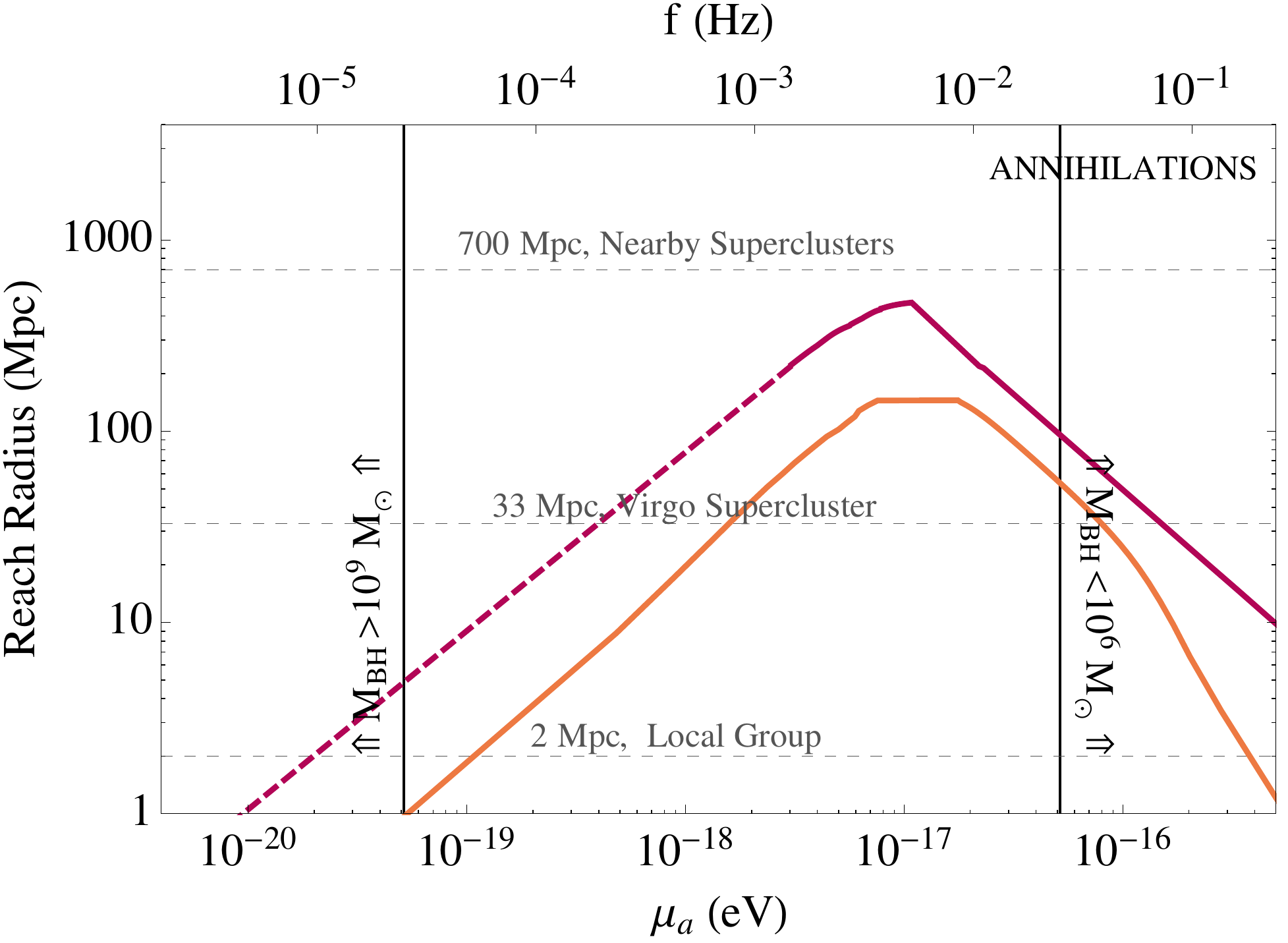}
    \vspace{-4mm}
    \caption{ AGIS (red/dark gray) and eLISA (orange/light gray) reach
      as a function of the axion mass for a black hole ``atom''
      currently undergoing $2p$ annihilations at the maximum rate for
      BH spin $a_{*} = 0.99$.  For $a_* = 0.9$ the reach is decreased
      by a constant factor $\sim 50$ as in fig.\ref{fig:treach}; for
      $a_*=0.7$ the reach decreases below $1\mpc$ for both
      experiments, making the signal unobservable for extragalactic sources.
    } \label{fig:areach_sm} \vspace{-5mm}
  \end{center}
\end{figure}

\begin{itemize}
\item Narrow: $M_{\rm{max}}  = 30\msun$
\item Intermediate: $M_{\rm{max}}  = 80\msun$
\item Wide: $M_{\rm{max}}  = 160\msun$
\end{itemize}

Within each band, we fix $ M_{\rm{min}} = 4.1\msun$ and vary $M_0$ from
$2.8$ to $7.9\msun$ with the central curves given by $M_0 =4.7\msun$.
Not much is known about the heavier
stellar BHs: the heaviest known stellar BH has mass
$32.7\pm 2.6 \msun$ \cite{silverman2008ic}, while theoretical modeling
of BH formation from a single star can accommodate maximum stellar BH
mass of $30$--$80\msun$ depending on the metallicity of the environment
\cite{belczynski2010maximum}.

\textit{Spin distribution}: The measured distribution is peaked at
high spins ($30\%$ above $0.8$) which we take as a realistic
estimate. We consider $90\%$ above $0.8$ as optimistic -- a high birth
spin is likely since the progenitor star has to lose a lot of angular
momentum to collapse to the small BH, and in the case a light boson is
present, some observed BHs would have spun down since their birth. We
take a flat initial spin distribution as pessimistic.

\textit{Formation rate}: Barring rare violent events (e.g. neutron
star-BH or BH-BH mergers, $< 10^{-2}$ per century in the Milky Way
\cite{Abadie:2010cf}), event probability is proportional to the BH
birth rate.  Core collapse supernovae rates are estimated to be
$1.9\pm 1.1$ per century \cite{Diehl:2006cf,Prantzos:2003ph}. The
fraction of supernovae that form black holes is estimated to be $15\%$
in metal-rich stars like the Sun \cite{heger2003massive} with a large
uncertainty. Based on average metallicities today, $20\pm 10\%$
\cite{heger2003massive,Zhang:2007nw} of supernovae form BHs. This
leads us to an optimistic value of $0.9$, realistic of $0.38$, and
pessimistic of $0.08$ BHs formed per century.

Violent BH formation may impede superradiant growth and delay the
signal until a more uniform accretion disk forms; our conclusions are
unaffected by delays less than a Gyr, and the unlikely case of delays
$>5$~Gyr would increase the signal by a factor of 3 due to higher star
formation rates \cite{Bouwens:2009qs}.

\textit{Distance distribution}: We assume the distance distribution of
BHs in the Milky Way is proportional to the stellar distribution
\cite{sale2010structure}.  Outside of our galaxy we scale the number
of BHs in our galaxy $N_{MW}$ by the blue luminosity distribution in
\cite{kopparapu2008host}, which asymptotes to $ N_{BH}= 0.0042\,
N_{MW}\, (r/\mathrm{Mpc})^3$ at distances $r>30\mpc$.

\vspace{-.5cm}
\SkipTocEntry\subsection{Supermassive Black Hole Distributions}

\textit{Mass and distance distribution}: Supermassive BHs are
generally understood to reach their mass through accretion, with most
BHs with mass $10^{6}\mathrm{-}10^{7}\msun$ and a tail extending to
$10^{10}\msun$. We use the distributions of
\cite{kelly2012mass,shankar2009self}, which give a total amount of
mass in SMBHs as $\rho_{BH} = (3.2\mathrm{-}5.4) \times 10^5\msun
\mpc^{-3} $, or about one $10^7\msun$ BH per MW-type galaxy; we
extrapolate down to $10^{5.5}\msun$ from $10^{6}\msun$. We scale this
distribution according to \cite{kopparapu2008host} at distances closer
than $30\mpc$.

\textit{Spin distribution}: The biggest uncertainty for event rates is
due to the unknown spin distribution of SMBHs. Some simulations of
thin disk accretion find that 70\% of SMBHs are maximally rotating
\cite{volonteri2005distribution}.  Black holes quickly spin up to
maximal spin, where maximal is less than $1$ due to counteracting
torques from either radiation emitted from the disk and absorbed by
the BH ($a_*^{\rm{max}} = 0.998$) \cite{thorne1974disk}, or magnetic
fields transporting angular momentum away from the BH
($a_*^{\rm{max}} = 0.93$) in simulations of thick disk models
\cite{gammie2004black}. We use 70\% of SMBHs with spins
$a_* \geq 0.93$ as an optimistic estimate. For a pessimistic estimate,
we use the chaotic accretion model \cite{King_2008} which gives low
spin $a_* =0.2\pm 0.2$; this leads to less than $10^{-2}$ events but
is also disfavored by measurements of many rapidly spinning BHs.
Merger dominated models give $a_* \sim 0.7$ \cite{Berti_2008}. For a
realistic estimate we use the hybrid model of \cite{Sesana_2014} where
70\% of SMBHs have spins $a_* \geq 0.7$ with 50\% above $ 0.9$ (for
BHs below $10^7\msun$ in mass, which are the ones contributing
dominantly to the signal).




\FloatBarrier

\SkipTocEntry\bibliography{sr}

\end{document}